\documentclass[%
reprint,
superscriptaddress,
nofootinbib,
amsmath,amssymb,
aps, prd,
bm
]{revtex4-2}

\usepackage{graphicx}%
\usepackage{dcolumn}%
\usepackage{bm}%
\usepackage{hyperref}
\usepackage[dvipsnames]{xcolor} %
\usepackage{braket}
\usepackage{todonotes}
\usepackage{MathUnicode}
\usepackage{newtxtext,newtxmath}
\usepackage{xspace} %
\usepackage{tabu}
\usepackage{multirow}
\usepackage{lineno}
\usepackage{svg}
\usepackage{tikz}
\usepackage{mathtools} %
\usepackage[dvipsnames]{xcolor}
\hypersetup{    
  colorlinks      = true,
  linkcolor       = {black},
  linkbordercolor = {white},
  citecolor       = {black},
  citebordercolor = {white},
  urlcolor        = {blue},
  urlbordercolor  = {white},
}%

\newcommand{\lensingcov}{\mathsf{C}_{\Delta\Sigma,jk}}

\defcitealias{costanzi_cosmological_2021}{C21}

\begin{document}

\defcitealias{costanziCosmologicalConstraintsY12021}{C21}
\newcommand{\Costanzi}{\citetalias{costanziCosmologicalConstraintsY12021}\xspace}

\newcommand{\avg}[1]{\left\langle{#1}\right\rangle}

\newcommand{\vecs}{\ln \boldsymbol{s}}
\newcommand{\vecalpha}{\boldsymbol{\alpha}}

\newcommand{\vecx}{\boldsymbol{x}}

\newcommand{\bsel}{b_\mathrm{sel}}

\newcommand{\beqa}{\begin{equation}\begin{aligned}}
\newcommand{\eeqa}{\end{aligned}\end{equation}}

\newcommand{\hiGpc}{h^{-1} \rm Gpc}
\newcommand{\hiMpc}{h^{-1} \rm Mpc}
\newcommand{\hikpc}{h^{-1} \rm kpc}
\newcommand{\hiMsun}{h^{-1} M_\odot}
\newcommand{\Msun}{M_\odot}

\newcommand{\Mvir}{M_{\rm vir}}
\newcommand{\OmegaM}{\Omega_{\rm M}}

\newcommand{\Mwl}{\Delta\Sigma}
\newcommand{\DS}{\Delta\Sigma}
\newcommand{\dsigma}{\mathbf{\Delta\Sigma}}
\newcommand{\rp}{r_{\rm p}}
\newcommand{\lambdabin}{\lambda_{\mathrm{bin}}}
\newcommand{\lambdamin}{\lambda_{\mathrm{min}}}
\newcommand{\lambdamax}{\lambda_{\mathrm{max}}}

\newcommand{\zetabin}{\zeta_{\mathrm{bin}}}
\newcommand{\zetamin}{\zeta_{\mathrm{min}}}
\newcommand{\zetamax}{\zeta_{\mathrm{max}}}

\newcommand{\xibin}{\xi_{\mathrm{bin}}}
\newcommand{\code}[1]{\textbf{\texttt{#1}}} %

\newcommand{\redmapper}{redMaPPer\xspace} %

\newcommand{\vecpi}{\boldsymbol{\pi}}
\newcommand{\vecmu}{\ln \boldsymbol{\mu}}
\newcommand{\alphaCalpha}{\boldsymbol{\alpha}^T \mathcal{C}^{-1} \boldsymbol{\alpha}} 
\newcommand{\alphaCspi}{\boldsymbol{\alpha}^T \mathcal{C}^{-1}(\vecs - \vecpi)} 
\newcommand{\sigmams}{\sigma_{m \mid \ln \mathbf{s}}}
\newcommand{\sigmamu}[1]{\sigma_{m|{#1}}}
\newcommand{\meanmu}[1]{\braket{m|{#1}}}
\newcommand{\normpdf}[2]{\exp\Bigg\{
    -\frac{1}{2}(\frac{#1}{#2})^2
    \Bigg\}}
\newcommand{\abacus}{\textsc{AbacusCosmos}\xspace}
\newcommand{\darkemu}{\textsc{Dark Emulator}\xspace}
\newcommand{\darkquest}{\textsc{Dark Quest}\xspace}

\newcommand{\hmpc}{$h^{-1}\text{Mpc}$ }
\newcommand{\hmsun}{$h^{-1} \mathrm{M}_{\odot}$ }

\newcommand{\mlambda}{\frac{\ln s - \pi_{\ln s}}{\alpha_{\ln s}}}
\newcommand{\mzeta}{\frac{\ln\zeta - \pi_\zeta}{\alpha_\zeta}}
\newcommand{\modelvec}{\mathbf{\Delta\Sigma}^{\mathrm{model}}_{jk} (\boldsymbol{\Theta})}
\newcommand{\unbiasvec}{\mathbf{\Delta\Sigma}^{\rho=0}_{jk}}
\newcommand{\datavec}{\mathbf{\Delta\Sigma}^{\mathrm{data}}_{jk}}
\newcommand{\noisevec}{\mathbf{\delta}^{\mathrm{model}}_{jk}(\Theta)}

\newcommand{\mpivotcostanzi}{\ln(3\times10^{14}M_\odot h^{-1})}
\newcommand{\growthcostanzi}{\ln \left(\frac{E(z)}{E(0.6)}\right)}

\newcommand{\lamob}{\lambda^{\rm ob}}
\newcommand{\sigmalam}{\sigma_\lambda}
\newcommand{\sigmaPoi}{\sigma_{\rm Poisson}}

\title[Optical $\times$ SZE clusters]{Forecasting the constraints on optical selection bias and projection effects of galaxy cluster lensing with multiwavelength data} 

\author{Conghao Zhou}
\affiliation{Physics Department, University of California, Santa Cruz, CA 95064, USA\\
Santa Cruz Institute for Particle Physics, Santa Cruz, CA 95064, USA}
\email{zhou.conghao@ucsc.edu}
\author{Hao-Yi Wu}
\affiliation{
Department of Physics, Boise State University, Boise, ID 83725, USA
}
\author{Andr\'es N. Salcedo
}
\affiliation{Steward Observatory, University of Arizona, 933 North Cherry Avenue, Tucson, AZ 85721, USA\\
Department of Physics, University of Arizona, 1118 East Fourth Street, Tucson, AZ 85721, USA}

\author{Sebastian Grandis}
\affiliation{Universität Innsbruck, Institut für Astro- und Teilchenphysik, Technikerstr. 25/8, 6020 Innsbruck, Austria}

\author{Tesla Jeltema}
\affiliation{Physics Department, University of California, Santa Cruz, CA 95064, USA\\
Santa Cruz Institute for Particle Physics, Santa Cruz, CA 95064, USA}

\author{Alexie Leauthaud}
\affiliation{Department of Astronomy and Astrophysics, University of California, Santa Cruz, 1156 High Street, Santa Cruz, CA 95064 USA}
\author{Matteo Costanzi}
\affiliation{Dipartimento di Fisica - Sezione di Astronomia, Universitá di Trieste, Via Tiepolo 11, 34131 Trieste, Italy\\
INAF-Osservatorio Astronomico di Trieste, Via G. B. Tiepolo 11, 34143 Trieste, Italy\\
IFPU, Institute for Fundamental Physics of the Universe, via Beirut 2, 34151 Trieste, Italy}
\author{Tomomi Sunayama}
\affiliation{Steward Observatory, University of Arizona, 933 North Cherry Avenue, Tucson, AZ 85721, USA}
\author{David H. Weinberg}
\affiliation{Center for Cosmology and AstroParticle Physics (CCAPP),
the Ohio State University, Columbus OH 43210, USA\\
Department of Astronomy, the Ohio State University, Columbus, OH 43210, USA}
\author{Tianyu Zhang}
\affiliation{Department of Statistics and Data Science, Carnegie Mellon University}
\author{Eduardo Rozo}
\affiliation{Department of Physics, University of Arizona, 1118 East Fourth Street, Tucson, AZ 85721, USA}
\author{Chun-Hao To}
\affiliation{Center for Cosmology and AstroParticle Physics (CCAPP),
the Ohio State University, Columbus OH 43210, USA\\
Department of Astronomy, the Ohio State University, Columbus, OH 43210, USA\\
Department of Physics, the Ohio State University, Columbus, OH 43210, USA}
\author{Sebastian Bocquet}
\affiliation{University Observatory, Faculty of Physics, Ludwig-Maximilians-Universität, Scheinerstr. 1, 81679 Munich, Germany}
\author{Tamas Varga}
\affiliation{Max Planck Institute for Extraterrestrial Physics, Gießenbachstr. 1, 85748 Garching, German\\
Excellence Cluster Origins, Boltzmannstr. 2, 85748 Garching, Germany\\
Universitäts-Sternwarte, Fakultät für Physik, Ludwig-Maximilians
Universität München, Scheinerstr. 1, 81679 München, Germany
}
\author{Matthew Kwiecien}
\affiliation{Physics Department, University of California, Santa Cruz, CA 95064, USA\\
Santa Cruz Institute for Particle Physics, Santa Cruz, CA 95064, USA}

\begin{abstract}
Galaxy clusters identified with optical imaging tend to suffer from projection effects, which impact richness (the number of member galaxies in a cluster) and lensing coherently.  Physically unassociated galaxies can be mistaken as cluster members due to the significant uncertainties in their line-of-sight distances, thereby changing the observed cluster richness; at the same time, projection effects alter the weak gravitational lensing signals of clusters, leading to a correlated scatter between richness and lensing at a given halo mass.  As a result, the lensing signals for optically selected clusters tend to be biased high.  This optical selection bias problem of cluster lensing is one of the key challenges in cluster cosmology.  Fortunately, recently available multiwavelength observations of clusters provide a solution.   We analyze a simulated data set mimicking the observed lensing of clusters identified by both optical photometry and gas properties, aiming to constrain this selection bias.  Assuming a \redmapper sample from the Dark Energy Survey with South Pole Telescope Sunyaev-Zeldovich effect observations, we find that an overlapping survey of 1300 deg$^2$, $0.2 < z < 0.65$, can constrain the average lensing bias to an accuracy of 5 \%.
This provides an exciting opportunity for directly constraining optical selection bias from observations.  We further show that our approach can remove the optical selection bias from the lensing signal, paving the way for future optical cluster cosmology analyses.
\end{abstract}

\maketitle

\section{Introduction}

The abundance of galaxy clusters as a function of cluster mass across cosmic time is a sensitive probe of cosmology, especially the growth of structure and cosmic acceleration [see, e.g.,~\cite{allenCosmologicalParametersObservations2011, weinbergObservationalProbesCosmic2013, hutererGrowthCosmicStructure2015} for reviews].  
Given that cluster mass is not directly observable, we rely on mass proxies observed across the electromagnetic spectrum, including optical richness (the weighted number of member galaxies in a cluster)\footnote{We focus on the richness defined by optical-band magnitudes and colors, which includes line-of-sight galaxies due to their significant redshift uncertainties. We refer to the number of galaxies within 3D halo radii as the halo occupation.} 
\citep{rykoffRedMaPPerAlgorithmSDSS2014}, 
X-ray properties (luminosity, temperature, gas mass) \citep{mantzObservedGrowthMassive2010, chiuCosmologicalConstraintsGalaxy2023}, 
and the Sunyaev--Zeldovich effect (SZE, inverse-Compton scattering of cosmic microwave background photons caused by the hot electrons in clusters) \cite{sunyaevSmallscaleFluctuationsRelic1970, saroConstraintsRichnessMass2015, saroOpticalSZEScalingRelations2017, bocquetClusterCosmologyConstraints2019, costanziCosmologicalConstraintsY12021}. Multiwavelength data sets allow us to cross-validate different cluster samples and robustly constrain the observable--mass relations [see, e.g., \citep{prattGalaxyClusterMass2019} for a review]. If their systematics can be controlled, clusters have the potential to be among the most powerful cosmological probes at low redshift.

The gravitational lensing signal of a cluster--- the tangential shearing of background galaxy images around the cluster---directly probes its mass [see, e.g., \cite{umetsuClustergalaxyWeakLensing2020} for a review].  However, individual cluster lensing signals tend to be dominated by the noise associated with background galaxy intrinsic shapes and are affected by the projected density fluctuations along the cluster's sightline \cite{beckerACCURACYWEAKLENSINGCLUSTER2011, wuCovarianceMatricesGalaxy2019}.  
To increase the signal-to-noise of cluster lensing, we often combine the signal of multiple clusters; for example, we choose clusters with similar optical richness and measure their mean stacked lensing signal \cite{johnstonCrosscorrelationWeakLensing2007, simetWeakLensingMeasurement2017, melchiorWeaklensingMassCalibration2017, murataConstraintsMassRichnessRelation2018, mcclintockDarkEnergySurvey2019}.  Calibrating the cluster observable--mass relation using the stacked lensing signal has become a standard procedure in cluster cosmology \cite{2023arXiv231012213B}.

Optical imaging identifies clusters down to lower masses than X-ray and SZE and tends to provide greater statistical power.  Forecasts have shown that optical clusters have a constraining power comparable to Stage IV CMB+SN+BAO+WL \citep{weinbergObservationalProbesCosmic2013}. 
However, this statistical power comes at a cost, as the richness--mass relation is shallower and has a larger scatter at a given mass than its X-ray and SZE counterparts. This makes optical selection particularly prone to projection effects; that is, richness can be boosted significantly by foreground and background galaxies due to the uncertainty in their line-of-sight distances \cite{cohnRedsequenceClusterFinding2007, nohGeometryFilamentaryEnvironment2011, buschAssemblyBiasSplashback2017, zuLevelClusterAssembly2017, sunayamaImpactProjectionEffects2020, wuOpticalSelectionBias2022, zhangEffectSelectionTale2022, zhangModellingGalaxyCluster2023}. These projection effects lead to one of the key systematics in optical cluster cosmology---the optical selection bias of stacked cluster lensing \citep{abbottDarkEnergySurvey2020, sunayamaImpactProjectionEffects2020, wuOpticalSelectionBias2022, sunayamaOpticalClusterCosmology2023}.  At a given cluster mass, the scatter in richness and lensing tend to be positively correlated; for example, a filament along the cluster sight-line can boost the richness and lensing signal simultaneously.
When we select a cluster sample above a given richness threshold, we inevitably select clusters with a boosted lensing signal relative to their mass. 
Therefore, correcting for the lensing bias induced by richness selection is equivalent to constraining the correlated scatter between richness and lensing \cite{whiteClusterGalaxyDynamics2010, evrardModelMultipropertyGalaxy2014}.
To realize the constraining power of optical clusters from wide-field imaging surveys \cite{2005astro.ph.10346T, 2009arXiv0912.0201L, 2011arXiv1110.3193L, dejongKiloDegreeSurvey2013, aiharaHyperSuprimeCamSSP2018, 2018arXiv180403628D}, it is imperative that we solve this projection effect problem.  
In this work, we explore the possibility of using multiwavelength cluster observables to constrain the correlated scatter between optical richness and lensing at a given mass.

The DES-Y1 cluster cosmology key project \cite{abbottDarkEnergySurvey2020} has identified the lensing bias associated with richness selection as one of the key systematics in mass calibration.  
\citet{sunayamaImpactProjectionEffects2020} have used mock cluster catalogs constructed using halo occupation distribution and counts-in-cylinders to study the impact of projection on cluster richness and lensing.  \citet{wuOpticalSelectionBias2022} have used the Buzzard simulations (synthetic catalogs for DES) \citep{deroseBuzzardFlockDark2019} to quantify the lensing bias associated with the \redmapper richness selection, finding a significant scale-dependent lensing bias across all richness bins.  They have also found that the counts-in-cylinders approach closely mimics the projection effects of \redmapper richness.  \citet{salcedoDarkEnergySurvey2023} have applied this forward modeling framework to DES-Y1 lensing and abundance data and showed it is consistent with {\it{Planck}} cosmology. \citet{zengSelfcalibratingOpticalGalaxy2023} have presented a mock analysis using the cross-correlation functions between clusters, galaxies, and lensing to constrain this selection bias.  In contrast, our current work aims to directly constrain the selection bias using multiwavelength observations. %

As a case study, we perform a mock analysis for the cluster lensing signal of a sample jointly selected by richness and SZE.  We develop a Monte Carlo-based modeling formalism: Starting from halo masses drawn from an analytical halo mass function, we simulate cluster observables with observable--mass distributions and model the correlated scatter between richness and lensing at a given mass.  We then predict the observed stacked lensing signals in richness and SZE bins.

In parallel, we develop mock catalogs for DES $\times$ SPT by assigning richness, lensing, and SZE to cluster-size halos in N-body simulations.  
Our mock catalog is constructed with a halo occupation distribution (HOD) framework, which models the number and spatial distributions of galaxies in halos as a function of halo mass \citep{berlindHaloOccupationDistribution2002, kravtsovDarkSideHalo2004, zhengTheoreticalModelsHalo2005}. 
We first simulate {\em all} red-sequence galaxies regardless of their host halo mass because they can potentially contribute to cluster richness through projection.  We then simulate the impact of projection on richness using counts-in-cylinders around massive halos \citep{costanziModelingProjectionEffects2019, sunayamaImpactProjectionEffects2020}.
We calculate the stacked cluster lensing signals using projected correlation functions between clusters and dark matter particles.  The counts-in-cylinder mock cluster finding and mock lensing self-consistently include the projected galaxies and matter along the cluster sightline. In addition, we assign SZE signals to halos using the best-fit scaling relation from \citet[][C21 hereafter]{costanziCosmologicalConstraintsY12021}, including intrinsic scatter and observational noise.  We expect that current X-ray and SZE mass proxies have a scatter nearly uncorrelated with lensing and richness \cite{mantzObservedGrowthMassive2010, saroConstraintsRichnessMass2015, saroOpticalSZEScalingRelations2017, bocquetClusterCosmologyConstraints2019, costanziCosmologicalConstraintsY12021, chiuCosmologicalConstraintsGalaxy2023}.

\begin{figure*}
    \includegraphics[width=\textwidth]{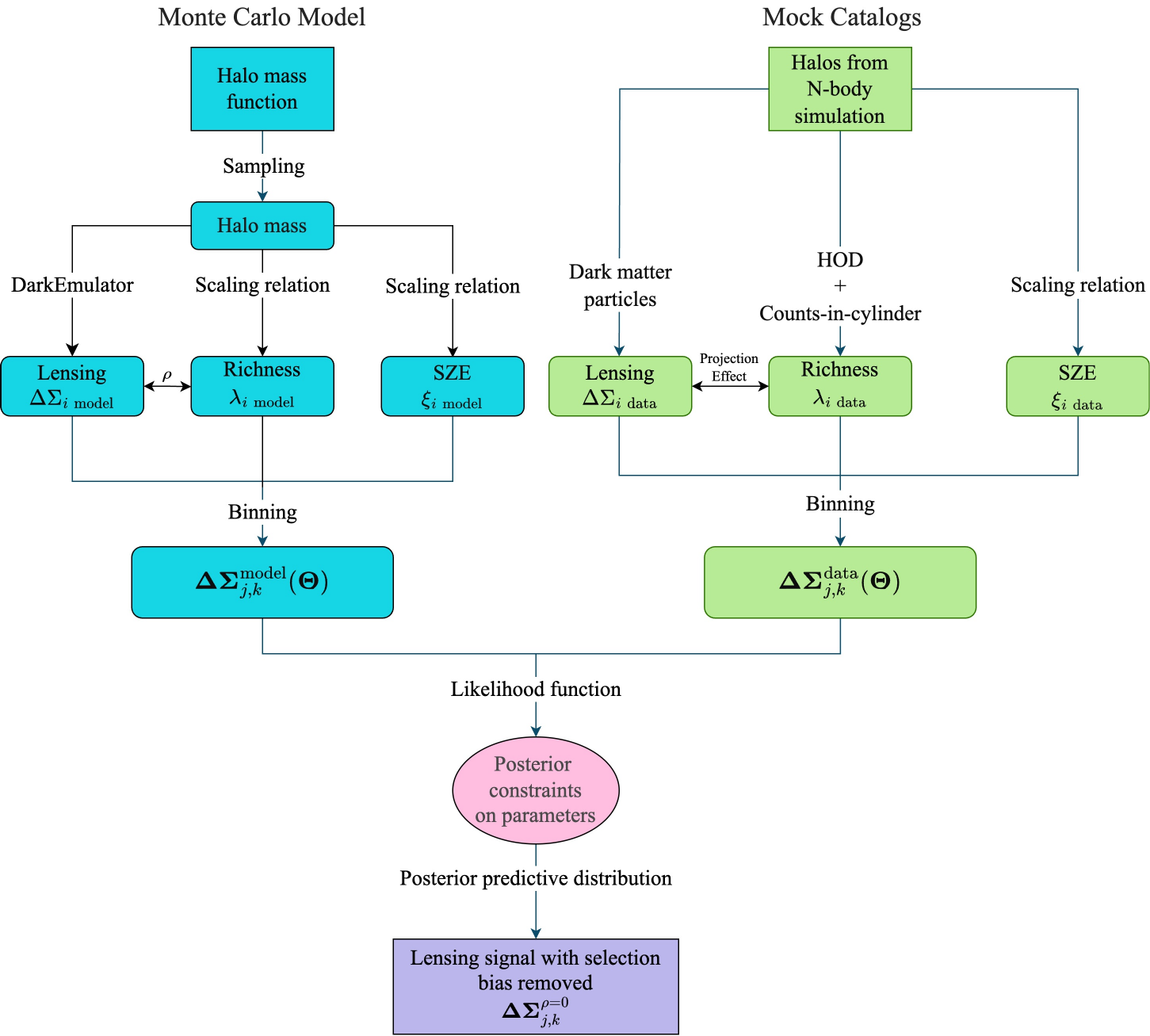}
    \caption{Our analysis pipeline. In the top left, in cyan, is the Monte Carlo model for $\Delta\Sigma$ that we introduce in Sec.~\ref{sec:montecarlo}. In the top right, in green, is the process for generating the mock data vector that we describe in Sec.~\ref{sec:mocks}. 
    We perform a likelihood analysis
    combining the model prediction and data vector and derive unbiased stacked lensing in richness and SZE bins in Sec.~\ref{sec:results}.
    } 
    \label{fig:flowchart} 
\end{figure*}

We then apply our modeling formalism to analyze these mock cluster catalogs.  We show that we can constrain the correlated scatter between richness and lensing by jointly selecting clusters using richness and SZE, thereby removing the lensing bias due to the richness selection.  A flowchart of our analysis is shown in Fig.~\ref{fig:flowchart}.

The next section presents our analytical formalism for multiwavelength cluster observables, focusing on the stacked lensing signal in the presence of correlated scatter.  Section~\ref{sec:montecarlo} presents our Monte Carlo-based model for predicting the lensing data vectors.  Section~\ref{sec:mocks} describes our mock catalogs that self-consistently include projection effects of richness and lensing. In Sec.~\ref{sec:results}, we present our mock likelihood analysis and demonstrate the constraining power of our approach.  We discuss our plan for applying our model to real data in Sec.~\ref{sec:discussion} and summarize in Sec.~\ref{sec:summary}.

\section{Analytical framework}
\label{sec:analytics}

In this section, we derive the stacked lensing signal for a given sample selection in the presence of correlated scatter. We assume logarithmic cluster observables follow a multivariate Gaussian distribution with a constant covariance matrix.  In Sec.~\ref{sec:montecarlo}, we will implement more flexible observable--mass distributions. 
Table~\ref{table:notations} lists our notations and definitions.

\begin{table*}
\caption{Notations of definitions in our analytic formalism.}
\label{table:notations}
\begin{tabular}{ll}
\hline
\multicolumn{2}{c}{Generic properties} \\\hline
$m \equiv\ln M$ & natural log of halo mass ($M_{500c}$ in $\hiMsun$, unless otherwise specified) \\
$s$, $\mu(m)$ & a generic cluster observable and its mean value at a given mass\\
$\alpha_s$, $\pi_s$, $\sigma_s$ & slope, intercept, and scatter of the $\ln s$--m relation \\
\hline
\multicolumn{2}{c}{Optical properties} \\\hline
$\lambda$ &  photometric richness defined by \redmapper  \\
$\rp$ & projected distance to cluster center (in comoving $\hiMpc$) \\
$\Delta\Sigma(\rp)$ & excess surface mass density; cluster weak lensing signal  \\
$\rho(\rp)$ & correlation coefficient between richness and lensing at a given mass\\
$\omega(\rp)$ & product of $\rho$ and $\sigma_{\Delta\Sigma}$ \\
\hline
\multicolumn{2}{c}{SZE properties} \\\hline
$\zeta$ & SZE intrinsic signal \\  
$\xi$   &  SZE observable (detection significance) \\
\hline
\multicolumn{2}{c}{Data vectors} \\\hline
$\datavec (\rp)$ & {\em observed} mean lensing signal in the $j$-th $\lambda$ bin and $k$-th $\xi$ bin \\
$\unbiasvec (\rp)$ & {\em predicted} mean lensing signal with the selection bias removed $(\rho = 0)$\\ 
$\bsel(\rp)$ &  ratio $\datavec/\unbiasvec$; the lensing bias due to richness selection \\
\hline
\end{tabular}
\end{table*}

\subsection{General model for multiple cluster observables}
\label{sec:general_analytical}

We follow the framework of multi-property cluster statistics in \citet{evrardModelMultipropertyGalaxy2014} with slight changes in notation. 
We denote the natural log halo mass as $m \equiv \ln {M}$ and the log cluster observable vector as
\begin{equation}
    \vecs \equiv (\ln s_1, \ldots, \ln s_n) ~ .
\end{equation}
These observables can be richness, X-ray temperature, SZE signal, etc. 
We assume $\vecs$ follows a multivariate Gaussian distribution 
\begin{align}
    P(\vecs | m)  
    &= \frac{\exp \left(-\frac{1}{2}(\vecs-\vecmu)^{\mathrm{T}} \mathsf{C}^{-1}(\vecs-\vecmu)\right)}{\sqrt{(2 \pi)^n|\mathsf{C}|}} ~ ,
\end{align}
with the mean vector
\begin{equation} \label{eq:meanscaling}
    \ln \mu_i(m) = \avg{\ln s_i|m} = \alpha_i m + \pi_i  ~ ,
\end{equation}
and the covariance matrix
\begin{equation}
    \mathsf{C}_{ij} = \avg{(\ln s_i-\ln \mu_i) (\ln s_j-\ln \mu_j)} ~.
\end{equation}
We assume that the covariance matrix is independent of mass and redshift.

\subsection{Mean lensing signal of a sample selected by richness: the origin of optical selection bias}

We first consider a specific example: the lensing signal, which is associated with the excess surface mass density $\Delta\Sigma(\rp)$, for a sample selected by richness $\lambda$.
Here, $\rp$ is the projected distance to the cluster center.
At a given halo mass, we assume that richness and lensing follow a bivariate Gaussian distribution
\begin{equation}
    P(\ln\Delta\Sigma, \ln \lambda|m) \sim \mathcal{N} \Big(\big\{\avg{\ln\Delta\Sigma|m},\avg{\ln\lambda|m}\big\}, \mathsf{C}\Big),
\end{equation}
with a covariance matrix 
\begin{equation}
    \mathsf{C}=\left(\begin{array}{cc}
\sigma_{\lambda}^2 & \rho \sigma_{\lambda} \sigma_{\Delta\Sigma} \\
\rho \sigma_{\lambda} \sigma_{\Delta\Sigma} & \sigma_{\Delta\Sigma}^2
\end{array}\right).
\end{equation}
At a given $m$ and $\lambda$, the probability density function (PDF) of lensing follows a Gaussian distribution
\begin{equation}
P(\ln\Delta\Sigma|\ln \lambda, m) = \frac{P(\ln\Delta\Sigma, \ln\lambda|m)}{P(\ln\lambda|m)} ~,
\end{equation}
     \\
with the mean given by
\begin{align}
   \avg{\ln \Delta\Sigma|\ln\lambda,m} &= \avg{\ln\Delta\Sigma|m} + \rho \sigma_{\Delta\Sigma} \frac{\ln\lambda - \avg{\ln\lambda|m}}{\sigma_{\lambda}}.
   \label{eq:bivariate_conditional}
\end{align}
This equation shows that, for a given halo, if $\ln\lambda$ is above the expectation from its mass, its lensing signal will also be above the expectation value from its mass. This shift is proportional to the product of $\rho$ and $\sigma_{\Delta\Sigma}$ and is inversely proportional to $\sigma_{\lambda}$.

The mean lensing signal in a richness bin is given by
\begin{equation} \label{eq:two_variables}
    \avg{\ln\Delta\Sigma|\lambdabin} = \int_{\lambdamin}^{\lambdamax} \avg{\ln\Delta\Sigma|\ln\lambda} P_{\lambdabin}(\ln \lambda) \; d \ln\lambda, 
\end{equation}
where we denote the normalized richness distribution in the bin as 
\begin{equation} \label{eq:normalised_pdf}
     P_{\lambdabin}(\ln\lambda) \equiv \frac{ P(\ln \lambda)}{\int_{\lambdamin}^{\lambdamax}  P(\ln \lambda) d \ln\lambda}.
\end{equation}
In Appendix~\ref{sec:lensing_lambda}, we show the analytic expressions for Eq.~(\ref{eq:two_variables}) assuming an exponential halo mass function.

\subsection{Mean lensing signal of a sample selected by richness and SZE} \label{sec:theory}

We now derive the mean lensing signal of clusters selected by both richness and SZE.  We denote the SZE intrinsic detection significance by $\zeta$.  For a given halo mass, we assume a multivariate Gaussian distribution for all log observables:
\begin{align}
&P(\ln\lambda, \ln \zeta, \ln\Delta\Sigma|m)
\nonumber\\
&\sim \mathcal{N} \Big(\big\{\avg{\ln\Delta\Sigma|m},\avg{\ln\lambda|m},\avg{\ln\zeta|m}\big\}, \mathsf{C}\Big).
\label{eq:3var_gaussian}
\end{align}
We further assume that, at a given mass, $\ln \lambda$ and $\ln \Delta\Sigma$ have a correlation coefficient $\rho$, while $\ln \zeta$ is uncorrelated with both.
\begin{equation}
\mathsf{C} = 
\begin{pmatrix} 
\sigma_{\Delta\Sigma}^2 && \rho \sigma_{\Delta\Sigma}\sigma_{\lambda} && 0 \\
\rho \sigma_{\Delta\Sigma}\sigma_{\lambda} && \sigma_{\lambda}^2 && 0 \\
0 && 0 && \sigma_{\zeta}^2
\end{pmatrix}  
~.
\label{eq:cov}
\end{equation}
The mean of $\ln \Delta\Sigma$ given $\ln \lambda$ and $\ln \zeta$ is given by
\begin{align}
    &\avg{\ln \Delta\Sigma|\ln \lambda, \ln \zeta} = \int_{-\infty}^{\infty} \ln \Delta\Sigma P(\ln \Delta\Sigma|\ln \lambda, \ln \zeta) \; d \ln \Delta\Sigma ~.
\end{align}

The mean $\Delta\Sigma$ in the $j$-th richness bin and the $k$-th SZE bin is given by
\begin{multline}
\modelvec 
\equiv
\avg{\ln\Delta\Sigma|\lambdabin, \zetabin}  
=
\int_{\lambdamin}^{\lambdamax}\int_{\zetamin}^{\zetamax} 
\\ \nolinenumbers
\left\langle \ln \Delta\Sigma \mid \ln \lambda, \ln \zeta\right\rangle
 P_{\lambdabin, \zetabin}(\ln \lambda, \ln \zeta) 
\; d\ln \zeta d\ln \lambda 
\label{eq:lensing_integral},
\end{multline}
where $P_{\lambdabin, \zetabin}(\ln \lambda, \ln \zeta)$ is the normalized joint distribution of log richness and SZE in the bin.  Eq.~(\ref{eq:lensing_integral}) is the generalization of Eq.~(\ref{eq:two_variables}) to three observables.  In Appendix~\ref{sec:lensing_richness_SZE}, we derive the analytical expression for Eq.~(\ref{eq:lensing_integral}) assuming an exponential mass function.

If richness and lensing have uncorrelated scatter ($\rho=0$), there is no lensing bias due to richness selection [see Eq.~(\ref{eq:bivariate_conditional})]; we denote the mean lensing signal under the zero-correlation assumption by $\unbiasvec$.
In mock catalogs, this quantity can be calculated by shuffling the richness of halos of the same mass, which washes out the correlated scatter.  It can also be calculated by finding the mass PDF of a richness-selected sample and calculating the lensing signal expected from this mass PDF.
We quantify the optical selection bias using
\begin{equation}
\bsel (\rp) = \frac{\datavec}{\unbiasvec} .
\label{eq:bsel}
\end{equation}
Under the Gaussian assumption, $\bsel$ depends on the product of the correlation coefficient $\rho$ and scatter $\sigma_{\Delta\Sigma}$ [see Eq.~(\ref{eq:bivariate_conditional})]. 
Therefore, we define the scaled correlation parameter as  
\begin{equation}
\omega(\rp) = \rho(\rp) \sigma_{\Delta\Sigma}(\rp) .
\label{eq:omega_def}
\end{equation}
In reality, $\omega$ also depends on mass as more massive halo tends to live in a more clustered environment. We do not include this mass dependence in this paper. We choose to model the $\Delta\Sigma$ profile instead of the weak lensing mass because, as we will see in Fig.~\ref{fig:lensing_abacus}, $\bsel$ exhibits a strong scale dependence, and the mass bias will depend on the scales used for mass calibration.     
Our goal is to recover the unbiased stacked lensing signal, which can be used for mass calibration or directly for cosmology analyses.

\section{Monte Carlo Implementation} \label{sec:montecarlo}

We build a Monte Carlo model to evaluate Eq.~(\ref{eq:lensing_integral}) with flexible observable--mass distributions.  We first draw halo masses from an analytical halo mass function and assign observables to these masses, assuming a set of scaling relations and the covariance.  We then calculate the mean $\Delta\Sigma(\rp)$ in richness and SZE bins to predict the observed lensing data vector.

\subsection{Mean richness, SZE, and lensing at a given mass} \label{sec:mean_richness_and_SZE}

We draw halo masses from the analytical mass function from \citet{tinkerHaloMassFunction2008} using inverse transform sampling.  We assume power law scaling relations for richness and the intrinsic SZE signal,
\begin{align}    
    \avg{\ln\lambda|m} &= \alpha_{\lambda} m + \pi_{\lambda} \\
    \avg{\ln\zeta|m} &= \alpha_{\zeta} m + \pi_{\zeta} . \
\end{align}
For the lensing profiles $\Delta\Sigma(\rp)$, we use the \darkemu \citep{nishimichiDarkQuestFast2019}, a Gaussian process emulator trained on the Dark Quest simulation ensemble covering 100 six-parameter $w$CDM cosmological models. We compare the $\Delta\Sigma$ from \darkemu with the $\Delta\Sigma$ from the \abacus simulations and find that the difference is $\lesssim$ 2\% between 0.1 and 100 $\hiMpc$.

\subsection{Observable--mass scatter}

For the scatter in the richness--mass relation, we adopt the mass-dependent richness scatter model from \citet{murataConstraintsMassRichnessRelation2018}:
$$
\sigma_{\lambda \mid m}=\sigma_{\lambda}+ q \left(m - m_{\text{pivot}} \right) ,
$$
in which 
$m_{\mathrm{pivot}}= \ln \left(3 \times 10^{14}~\hiMsun \right )$. We assume this scatter includes both intrinsic and observational scatter.

For a given halo mass and richness, we draw a lensing signal based on the conditional probability distribution to account for the correlated scatter between lensing and richness
\begin{multline}
    P(\ln \Delta\Sigma \mid m, \ln \lambda) \\ \nolinenumbers
    \sim \mathcal{N}\left(\langle \ln \Delta\Sigma \mid m \rangle + \frac{\sigma_{\Delta\Sigma}}{\sigma_{\lambda \mid m}}(\ln \lambda - \langle \ln\lambda \mid m \rangle),\left(1- \rho^2 \right) \sigma_{\Delta\Sigma}^2  \right) ~.
\end{multline}
The scatter of $\Delta\Sigma$ is from the individual lensing profiles in the \textsc{Cardinal} simulation \citep{toBuzzardCardinalImproved2023a}.  We describe the details in Appendix~\ref{sec:dsigma_scatter}.

For the observed SZE signal, we first add a Gaussian scatter to $\avg{\ln \zeta|m}$ to obtain the intrinsic signal $\zeta$. We then draw the observed signal $\xi$ based on this PDF from \citet{vanderlindeGalaxyClustersSelected2010}:
\begin{equation}
P (\xi | \zeta) = \mathcal{N}\left(
\sqrt{\zeta^2 + 3}, 1
\right) ~ .
\label{eq:xi}
\end{equation}

\subsection{Stacked lensing signal binned by richness and SZE}

For a given richness and SZE bin, the model prediction of log lensing is given by
\begin{equation}
\avg{\ln\Delta\Sigma|\lambda_{\mathrm{bin}}, \xi_{\mathrm{bin}}}_{\mathrm{model}} = \frac{1}{N} \sum_i \mathbf{1}_{\lambda_\mathrm{bin}}(\lambda_i) \mathbf{1}_{\xi_\mathrm{bin}}(\xi_i) \ln\Delta\Sigma_i  ,
\label{eq:lensing_montecarlo}
\end{equation}
where the index $i$ runs through all clusters, and
\begin{equation}
    N = \sum_i \mathbf{1}_{\lambda_\mathrm{bin}}(\lambda_i) \mathbf{1}_{\xi_\mathrm{bin}}(\xi_i)  ,
\end{equation}
is the number of halos in each bin.  Here $\mathbf{1}_A(x)$ is the indicator function defined as 
\begin{equation}
    \mathbf{1}_A(x):= \begin{cases}1 & \text { if } x \in A \\ 0 & \text { if } x \notin A\end{cases} .
\end{equation}
Eq.~(\ref{eq:lensing_montecarlo}) is a Monte Carlo estimator for Eq.~(\ref{eq:lensing_integral}).

\section{Mock data vectors}
\label{sec:mocks}

To validate our analysis pipeline, we generate mock catalogs with realistic projection effects.  
We assign galaxies to dark matter halos in N-body simulations using an HOD model, and we use projected galaxies and dark matter particles to simulate the richness and the lensing signal.  We use SZE--mass scaling relation to assign an SZE signal to each halo.

\subsection{Simulating cluster observables}

{\bf N-body simulations.}
We use the 20 boxes of 1100 \hmpc from the \abacus N-body simulation suite \citep{garrisonAbacusCosmosSuite2018} with a flat {\em Planck} $\Lambda$CDM cosmology \citep{adePlanck2015Results2016}: $\OmegaM$ = 0.314, $h$ = 0.673, $\sigma_8$ = 0.83, $n_{\rm s}$ = 0.9652, $\Omega_{\rm b}$ = 0.049.  We fix the cosmological parameters to these values throughout this paper.  The simulation has a mass resolution $4 \times 10^{10}~\hiMsun$, which allows us to use halos down to mass $10^{12}~\hiMsun$.  We use the central halos from the {\sc Rockstar} \citep{behrooziROCKSTARPHASESPACETEMPORAL2012} halo catalogs. Unless otherwise stated, we use $M_{500c}$, i.e., the halo mass enclosed by a spherical overdensity of 500 times the critical density of the universe.  When simulating galaxies in clusters, we follow the convention in the HOD literature and use $M_{200m}$ (an overdensity of 200 times the mean density of the universe).

{\bf SZE signals.}
We simulate the intrinsic SZE signal $\zeta$ using the best-fit scaling relation from \Costanzi, and we simulate the observed SZE significance $\xi$ based on Eq.~(\ref{eq:xi}).
We list the parameter values in Table~\ref{table:prior} and describe the conversion from \Costanzi in Appendix~\ref{sec:scaling_relation}.

{\bf Richness.}
We wish to create mock cluster catalogs with realistic projection effects.  First, we use an HOD model to simulate all galaxies with color--magnitude consistent with being a red-sequence cluster member regardless of their host halo mass.  These galaxies could be included as a member even if they are outside the cluster but are along the cluster sightline.
We then use counts-in-cylinders to calculate the cluster richness in the presence of projection effects 
\cite{sunayamaImpactProjectionEffects2020, costanziCosmologicalConstraintsY12021, wuOpticalSelectionBias2022, zengSelfcalibratingOpticalGalaxy2023, 2023arXiv231003944S}.  We emphasize that the HOD occupation is distinct from the cluster richness; occupation refers to the number of galaxies within the 3D virial radius of a halo, while richness refers to the number of member galaxies as determined by the photometric cluster finder.

To simulate red-sequence galaxies, we assume that all halos with $M_{\rm 200m} \ge M_{\rm min}$ host such a central galaxy
\begin{equation}
\avg{N_{\rm cen} | M_{\rm 200m}} = 
H(M_{\rm 200m} - M_{\rm min}) ~ .
\end{equation}
in which $H$ is the Heaviside step function. The number of satellite galaxies is drawn from a Poisson distribution with the mean
\begin{equation}
\avg{N_{\rm sat} | M_{\rm 200m}}
= \avg{N_{\rm cen} | M_{\rm h}} \left(
\frac{M_{\rm h}-M_0}{M_1}
\right)^{\alpha} ~.
\end{equation}
We adopt the HOD parameters used in \citet{sunayamaImpactProjectionEffects2020},
$\alpha = 1$,
$M_{\rm min} = 10^{12} \hiMsun$,
$M_{\rm 0} = 10^{11.7} \hiMsun$, and
$M_{\rm 1} = 10^{11.9} \hiMsun$.  
To estimate the richness, we count the number of galaxies in a cylinder of $\pm$ 30 $\hiMpc$ along the line-of-sight \citep{wuOpticalSelectionBias2022}. The aperture of the cylinder is iteratively determined using $R_\lambda = 1 (\lambda / 100)^{0.2} ~ {\rm physical}~\hiMpc$. We also mimic the percolation process of \redmapper: if a galaxy resides in the cylinders of multiple halos, it is only counted as the satellite of the most massive halo.  In \citet{zengSelfcalibratingOpticalGalaxy2023}, we have shown that the HOD+counts-in-cylinders approach produces number counts vs.~richness consistent with DES-Y1 clusters.

To facilitate the comparison with the power-law model used in Sec.~\ref{sec:montecarlo}, we fit a power-law richness--mass relation assuming a log-normal model using the maximum likelihood estimator (see Appendix~\ref{sec:mle} for details).  Given that our richness values are derived from counts-in-cylinders and do not necessarily follow a log-normal distribution, the fiducial values for the richness--mass relation are only approximate.

{\bf Stacked lensing signal.}
We compute the projected cross-correlation function between clusters and dark matter particles using {\sc corrfunc} \citep{sinhaCORRFUNCSuiteBlazing2020}
\begin{align}
\begin{aligned} 
w_{\rm p,cm}(\rp) &= 2 \int_0^{\Pi_\mathrm{max}} \xi_{\rm cm} \left(\rp, \pi \right) d \pi.
\end{aligned}
\end{align}
We then convert $w_{\rm p,cm}$ to the excess surface mass density
\begin{align}
\begin{aligned}
\Delta\Sigma (\rp) &= \Omega_m \rho_{\mathrm{crit}}  \left[ \frac{2}{\rp^2} \int_{0}^{\rp} r' w_{\rm p,cm}(r') dr' - w_{\rm p,cm}(\rp) \right].
\end{aligned}
\end{align}
We use a $\Pi_{\rm max} = 100~\hiMpc$, and 15 logarithmically spaced $\rp$ bins between 0.1 and 100 $\hiMpc$. We only use lensing signals with $\rp \geq 0.3 ~\hiMpc$, mimicking the radial range of DES lensing analyses \cite{mcclintockDarkEnergySurvey2019}. Throughout this work, unless otherwise specified, we use comoving distances.

{\bf Lensing covariance matrices.}  Given the relatively low source number density, our lensing covariance matrices are dominated by shape noise and are diagonal \cite{wuCovarianceMatricesGalaxy2019}
\begin{equation} 
\label{eq:shape_noise}
\mathrm{Var}\left(\Delta\Sigma(\rp)\right) = \Sigma_\mathrm{crit}^2 \frac{\sigma_\gamma^2}{n_\mathrm{src} N_\mathrm{cl} A(\rp)}~.
\end{equation}
Here $A(\rp)$ is the projected area of the radial bin.  For the cluster number $N_\mathrm{cl}$, we use the number of clusters expected from a 1300 deg$^2$ survey (see Table~\ref{table:counts}). For the critical surface density $\Sigma_\mathrm{crit}$, we assume a lens redshift $z_{\rm lens}=0.3$, a source redshift $z_{\rm src} = 0.75$.  We assume the noise level of the DES-Y1 \textsc{metacalibration} shape catalog: $\sigma_\gamma = \sigma_e / \sqrt{2} = 0.19 $, and a source density of $n_\mathrm{src} = 6.3$ galaxies per square arcmin \cite{zuntzDarkEnergySurvey2018}.

\begin{figure}
    \includegraphics[width=\columnwidth]{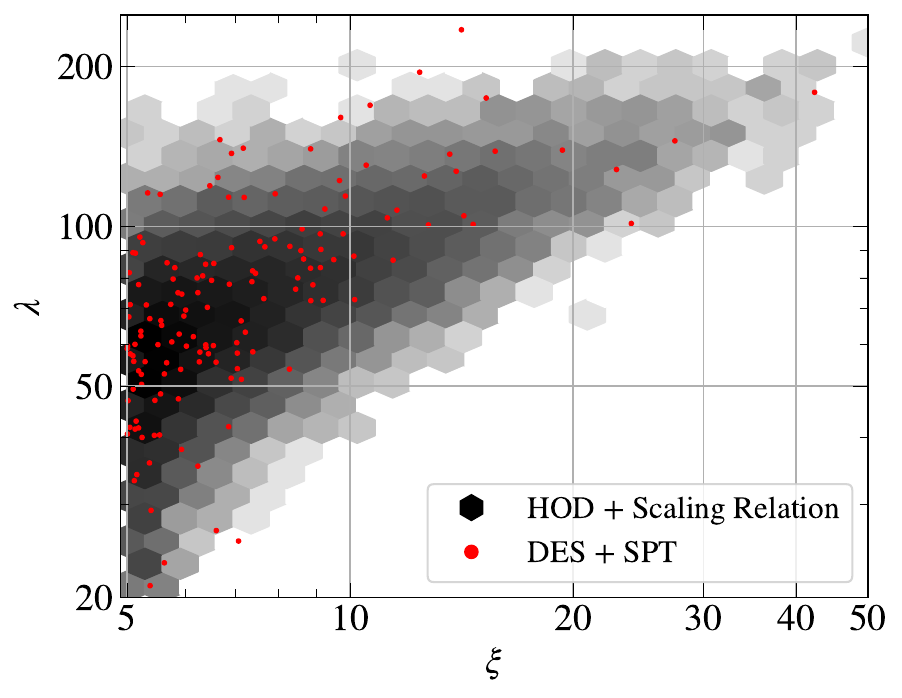}
    \caption{Richness vs.~SZE from our mock catalogs (hexagonal binning), compared with the DES$\times$SPT data from \protect\Costanzi (red points).
    To simulate cluster richness, we assign galaxies to halos in \abacus and count galaxies in cylinders along the line of sight.  The SZE signal is based on the SPT scaling relation and instrumental noise.  The agreement between our mock catalogs and observations indicates that our mock catalogs are realistic. }
\label{fig:lambda_xi}
\end{figure}

{\bf Comparison with observations.} Figure~\ref{fig:lambda_xi} shows the $\lambda$ and $\xi$ from our 20 mock catalogs.  The grayscale of the hexagonal binning is proportional to the log number density.   The red points are the 129 clusters from \Costanzi.  This sample is from the 1300 deg$^2$ overlapping area between the SPT-SZ sample with DES-Y1 \redmapper, with $0.2 < z < 0.65$, and it is derived by matching the positions and redshifts of clusters (see \Costanzi for details).  Our simulated richness and SZE signal agree well with observations.  This gives us the confidence to use these mock catalogs to test our analysis pipeline.  As we will discuss in Sec.~\ref{sec:discussion}, future improvements of the mock catalogs include incorporating projection effects of the SZE signal and the miscentering of the \redmapper clusters.

\subsection{Simulated data vector: lensing signal selected by both richness and SZE}

We first split clusters into two richness bins, [20, 50) and [50, $\infty$). For a given richness bin, we further split the clusters into SPT-detected ($\xi \ge 5$) and SPT-undetected ($\xi < 5$) samples. We denote the mock lensing signal in $j$-th richness and $k$-th SZE bin by $\datavec$. For a 1300 $\mathrm{deg}^2$ sky coverage, we expect $\approx 24~(4600)$ clusters with (without) SPT detection in the richness [20, 50) bin and $\approx 125~(356)$ clusters with (without) SPT detection in the richness [50, $\infty$) bin (see Table~\ref{table:counts}).

\begin{figure*}
   \includegraphics[width=\textwidth]{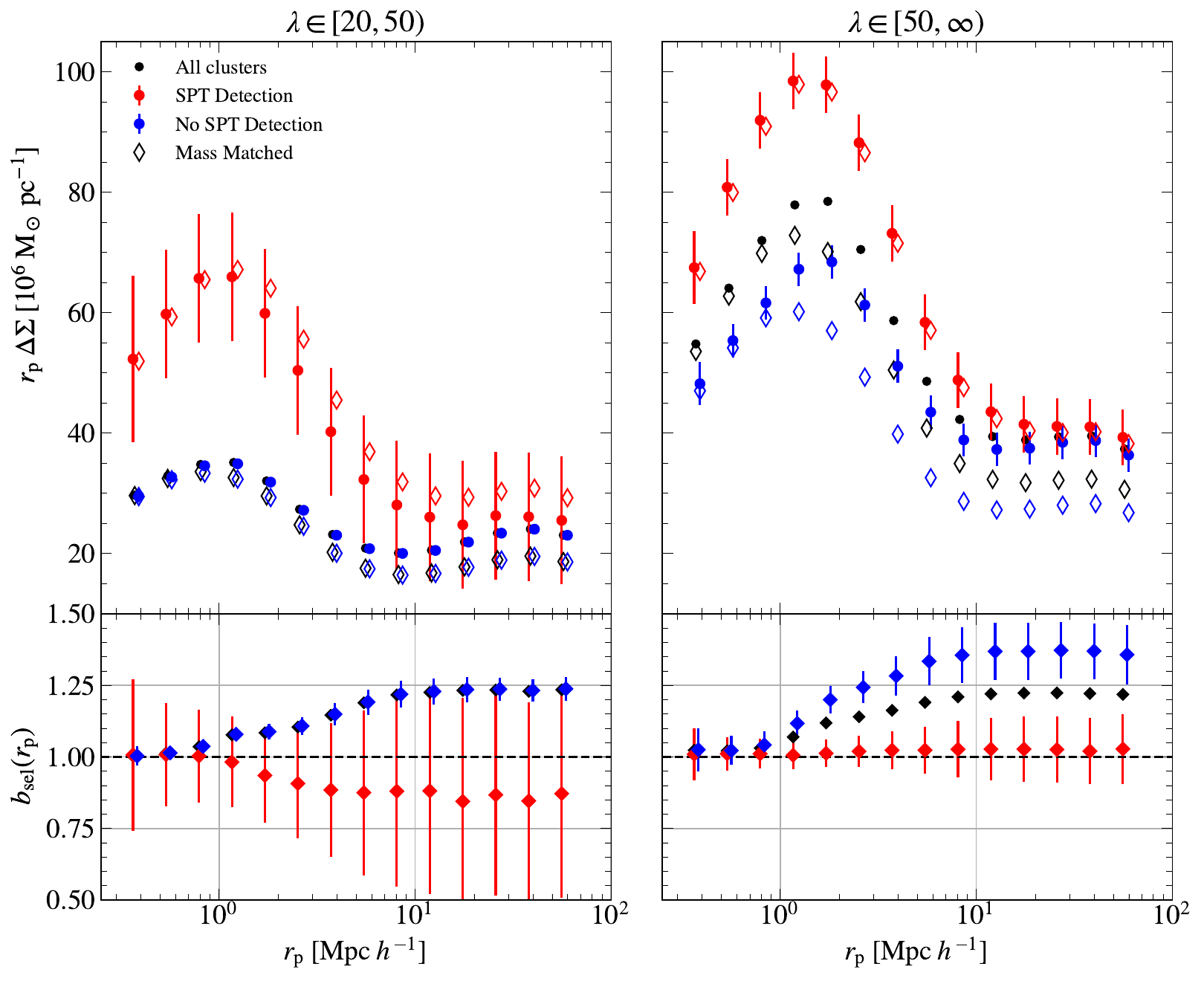}
    \caption{
    Lensing signals calculated from our mock catalogs. The left and the right panels are $\lambda \in [20,50)$ and $\lambda \in [50, \infty)$ bins.
    In each panel, the black markers correspond to all clusters, which are further split into SPT-detected (red) and SPT-undetected (blue). The error bars are estimated with Eq.~(\ref{eq:shape_noise}). Small horizontal offsets are added for discernability. The round points are the lensing signals of each sample, and the diamonds are the lensing signals of the mass-matched sample.  The bottom panels show the ratios between the round points and diamonds, which quantify the selection bias $\bsel(\rp)$. The clusters with no SPT detection (blue) tend to be more strongly biased than those detected by SPT (red). The low-richness clusters with SPT detection exhibit $\bsel < 1$ because they have richness and lensing signals lower than expected from their mean mass. 
    }
\label{fig:lensing_abacus}
\end{figure*}

Figure~\ref{fig:lensing_abacus} shows the lensing signal of our simulated samples and their corresponding selection bias, averaging over 20 mock catalogs.  The left-hand panels correspond to the low-richness bin, and the right-hand panels correspond to the high-richness bin.  In a given richness bin, we show all clusters (black), SPT-detected (red), and SPT-undetected (blue).  In the bottom left panel, the blue and black points overlap almost completely because most low-richness clusters are undetected by SPT.

The diamonds show the lensing signal {\em expected} without selection bias, $\unbiasvec$.  To derive this quantity, we calculate the mass PDF of a given sample and randomly draw halos from the full N-body simulation to match this mass PDF.  We draw with replacement a random sample that is ten times the size of the data to reduce the noise. In the bottom panel, we plot the ratio between the round points and the diamonds.  This ratio is the selection bias $\bsel$ we wish to infer from the likelihood analysis [see Eq.~(\ref{eq:bsel})].  We can see that the selection bias is scale-dependent and is consistent with the finding in \citet{sunayamaImpactProjectionEffects2020}.  The bias is negligible at 1-halo scales ($\rp<1~\hiMpc$) for all samples.  For all clusters in a richness bin (black), the selection bias $\bsel \geq 1$ at 2-halo scales.

In the bottom right panel, we can see that high-richness, SPT-undetected clusters (blue) tend to have the highest selection bias. These clusters tend to have low mass (thus no SPT detection) but have their richness and lensing significantly boosted by projection effects.  On the other hand, the SPT-detected clusters (red) have a nearly unbiased lensing signal.  In the bottom left panel, most clusters have no SPT detection because they tend to have low mass.  For the small fraction (0.7 \%) of SPT-detected clusters (red), they tend to have $\bsel < 1$.  These systems are likely massive (thus with SPT detection) but have downward-scattered richness and lensing signal.

\begin{table}
\caption{Cluster counts expected from a survey of 1300 deg$^2$, $0.2 \leq  z \leq 0.65$, derived from DES-Y1 \redmapper catalog and the SPT-SZ catalog from \protect\Costanzi.}
\label{table:counts}
\begin{tabular}{ccc}
\hline
& $\lambda \in [20,50)$ & $\lambda \in [50,\infty)$ \\ \hline
SPT detected ($\xi\ge 5$)  & 24  & 125   \\
SPT undetected ($\xi < 5$) & 4600& 356 \\ \hline
\end{tabular}
\end{table}
\begin{table*}
\caption{Free parameters and their priors in our MCMC. For the scaling relation parameters, we assume the priors at the level of \protect\Costanzi and \protect\citet{murataConstraintsMassRichnessRelation2018}. 
}
\label{table:prior}
\begin{tabular}{llllll}
\hline
Parameter        & Description & Prior & Prior Source & Fiducial Value \\ \hline
$\alpha_{\lambda}$ &    Slope of richness--mass relation & $\mathcal{U}(0, 160)$  & Flat prior & 0.84\\
$\pi_{\lambda}$    &    Intercept of the richness--mass relation & $\mathcal{U}(-100, 0)$  & Flat prior & -24.11\\
$\sigma_{\lambda}$ &    Intrinsic scatter of the richness--mass relation  & $\mathcal{N}(0.22, 0.064)$  &  MLE center + Costanzi width & 0.22\\
$q$ &    Mass dependence of the richness--mass relation& $\mathcal{N}(-0.10, 0.08)$ & MLE center + Murata width & -0.10\\
$\alpha_{\zeta}$   &    Slope of $\zeta$--mass relation    & $\mathcal{N}(1.53,0.14)$ &Costanzi center and width & 1.53\\
$\pi_{\zeta}$      &    Intercept of $\zeta$--mass relation& $\mathcal{N}(-49.44,4.67)$ &Costanzi center and width &-49.44\\
$\sigma_{\zeta}$   &    Scatter of $\zeta$--mass relation &  $\mathcal{N}(0.17, 0.085)$ & Costanzi center and width &0.17\\ \hline
$\omega_1$           &     Mean $\rho \sigma_{\Delta\Sigma}$ in 1-halo     &  $\mathcal{U}(-\sigma_{\Delta\Sigma_1}, \sigma_{\Delta\Sigma_1}$) & Flat prior \\ 
$\omega_2$           &     Mean $\rho \sigma_{\Delta\Sigma}$ in 2-halo   &  $\mathcal{U}(-\sigma_{\Delta\Sigma_2}, \sigma_{\Delta\Sigma_2}$) & Flat prior \\ 
\hline
\end{tabular}
\end{table*}

\begin{figure*}
    \includegraphics[width=\textwidth]{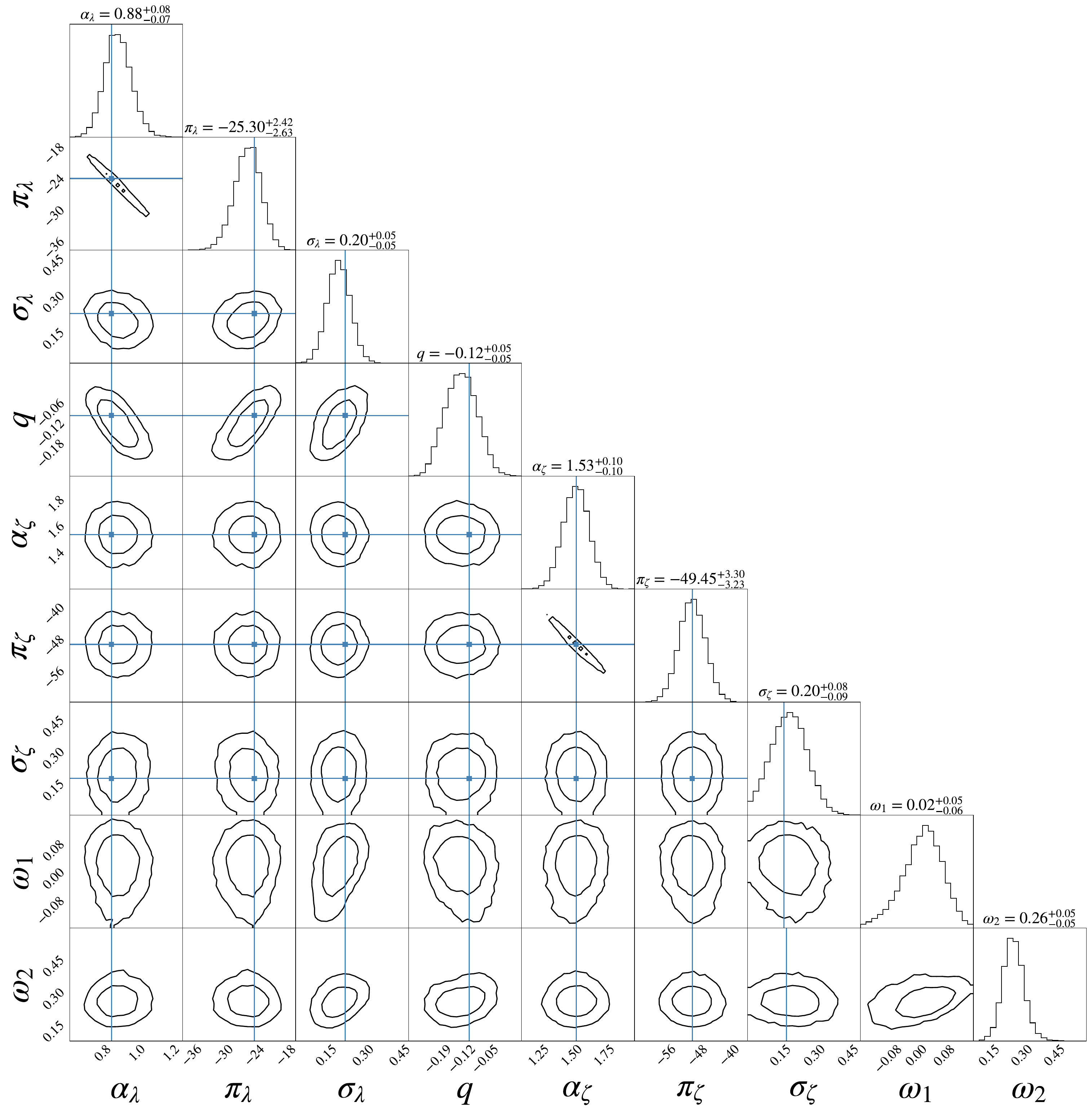}
    \caption{
    Posterior distribution of the full parameter space ($68\%$ and $95\%$ intervals).  The SZE parameters are dominated by the prior.  The blue vertical lines show the fiducial values from the simulation. We recover the fiducial scaling relation and constrain the $\omega$ parameter [Eq.~(\ref{eq:omega})] in the 1-halo and 2-halo to be $0.02 \pm 0.05$ and $0.26 \pm 0.05$.
    }
    \label{fig:posterior}
\end{figure*}

\begin{figure*}
    \includegraphics[width=\textwidth]{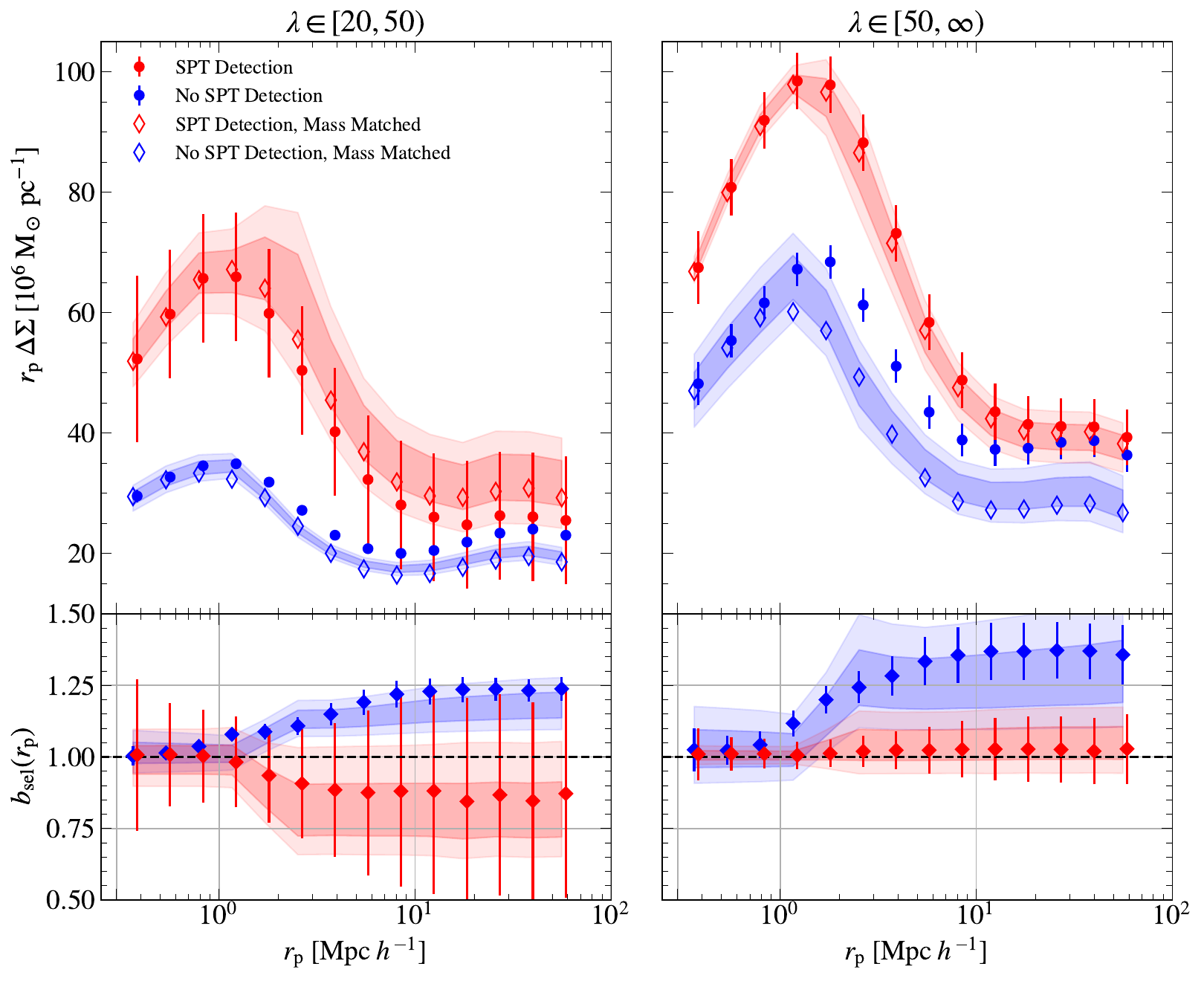}
    \caption{
    Lensing selection bias derived from our MCMC posterior distribution.  The data points are the same as in Fig.~\ref{fig:lensing_abacus}.  In the top panel, the bands show the unbiased lensing signals ($\unbiasvec$) derived from the posterior distribution (see Sec.~\ref{sec:results}), which are consistent with the diamonds derived from the halo mass PDF. Our posterior predictions correctly recover the unbiased lensing signals.  In the bottom panel, we show the lensing bias $\bsel(\rp)$ inferred from the posterior distribution. 
    } 
    \label{fig:recovered_dsigma} 
\end{figure*}

\section{Inferring the lensing selection bias} 
\label{sec:results}

In this section, we use our model (described in Sec.~\ref{sec:montecarlo}) to analyze the mock data vector (described in Sec.~\ref{sec:mocks}). Our goal is to infer the lensing selection bias from the data vector and recover the unbiased stacked mean lensing signal.

\subsection{Parameters and priors}

Given that the optical selection bias is scale-dependent and has constant values at small and large scales (Fig.~\ref{fig:lensing_abacus}), we assume that the scaled correlation parameter [Eq.~(\ref{eq:omega_def})] is given by 
\begin{equation}
    \omega(\rp) = 
    \begin{cases}
        \omega_1 & \text{if } \rp \in (0, r_1] \\
        \omega_1 + \frac{\omega_2-\omega_1}{r_2 - r_1}(\rp - r_1) & \text{if } \rp \in (r_1, r_2]\\
        \omega_2 & \text{if } \rp \in (r_2, r_3]
    \end{cases},
\label{eq:omega}
\end{equation}
where
$r_{1}=1 ~\hiMpc$,
$r_{2}=1.5~\hiMpc$,
and $r_{3}=100~\hiMpc$.
We have considered using each radial bin independently; however, a single radial bin does not provide enough signal-to-noise to constrain $\omega$. Therefore, we choose this asymptotic functional form.

The free parameters of our model are 
\begin{equation}
\boldsymbol{\Theta} = \{\alpha_{\lambda}, \pi_{\lambda}, \sigma_{\lambda}, q, \alpha_\zeta, \pi_\zeta, \sigma_\zeta, \omega_1, \omega_2\}.
\end{equation}
Their fiducial values and priors are summarized in Table~\ref{table:prior}.

Our likelihood analysis uses only the lensing information, and we need priors on the scaling relation parameters from cluster abundances.  For the SZE scaling relation and scatter, we use the constraints from \Costanzi, which are based on the combination of SPT-SZ externally-calibrated scaling relation and DES-Y1 number counts (the second column of Table IV in \Costanzi).
We note that the DES lensing signal is not used in their analysis; therefore, by using their constraints, we are not double-counting the lensing information.

For the richness scaling relation parameters, we assume no prior information. We need an extra prior on the mass-dependent richness--mass scatter, described by $q$.  We use the abundance-only constraints from \citet{murataConstraintsMassRichnessRelation2018}, which is based on the SDSS clusters ($\approx$8000 clusters in $\approx$10000 deg$^2$, $0.1<z<0.33$); see the fourth column in their Table 3.  We do not use their lensing+abundance constraints to avoid double-counting the lensing information.
In the future, we plan to carry out a comprehensive analysis incorporating all number counts and lensing binned by richness and SZE signals, as well as priors from follow-up observations.

\subsection{Inference scheme}

We define $\modelvec$ as the mean stacked lensing signal in the $j$-th richness bin and the $k$-th SZE bin.   Our log-likelihood function is given by $\mathsf{C}$ $\lensingcov$
\begin{align}
   &  \ln \mathcal{L} = \sum_{j,k} \ln \mathcal{L} \left(\datavec \mathsf{} \mid \boldsymbol{\Theta}  \right) = \\ 
   & \sum_{j,k}-\frac{1}{2} \left(\modelvec - \datavec \right)^T \lensingcov^{-1} \left(\modelvec - \datavec\right) ~,
\end{align} 
where $\lensingcov$ is the lensing covariance matrix described in Sec.~\ref{sec:mocks}.

We use \textsc{emcee} \citep{foreman-mackeyEmceeMCMCHammer2013}, a Python package based on the affine-invariant ensemble sampler \cite{goodmanEnsembleSamplersAffine2010}. We used 1600 walkers traversing 10000 steps in the 9-dimensional parameter space. We burn in 1000 samples, which is much greater than the autocorrelation time $\sim$ 150. The posterior distribution of the full parameter space is shown in Fig.~\ref{fig:posterior}. 
The SZE scaling relation parameters are prior dominated. The small-scale correlation $\omega_1$ is consistent with zero, and the large-scale correlation $\omega_2$ is significantly different from zero.  For each parameter, we mark the fiducial values with blue vertical and horizontal lines.

\subsection{Recovering the unbiased stacked lensing signals}

From the posterior distribution of the parameters, we can predict the lensing signal \textit{in the absence} of the correlated scatter between richness and lensing, which allows us to remove the effect of selection bias.  For a given set of parameters $\boldsymbol{\Theta}$ from the posterior, we calculate $\unbiasvec$, the stacked unbiased lensing signal in the $j$-th richness bin and $k$-th SZE bin by setting $\rho=0$.
We draw 10\% points from the posterior to calculate this quantity.
The selection bias $ \bsel(\rp | \boldsymbol{\Theta})$ is then calculated using Eq.~(\ref{eq:bsel}).

In Fig.~\ref{fig:recovered_dsigma}, we again show the mock lensing data (round points for $\datavec$ and diamonds for $\unbiasvec$) from Fig.~\ref{fig:lensing_abacus}.  
The color bands correspond to the 68\% and 95\% intervals of the posterior prediction.   As shown by the overlapping between contours and diamonds, our posterior prediction recovers the unbiased lensing within 1$\sigma$ levels.  
The posterior predictions have kinks at $\approx2~\hiMpc$, which correspond to the transition regions of our parameterization in Eq.~(\ref{eq:omega}).

Our results show we can derive the optical selection bias and recover the unbiased mass by fitting the lensing signals jointly selected by richness and SZE.
With a few modifications (discussed in Sec.~\ref{sec:discussion}), our likelihood analysis is readily applicable to real data.  
This opens up an exciting opportunity for directly inferring and removing optical selection bias from observational data.

\section{Discussion} \label{sec:discussion}

We first discuss the possible impact of projection on SZE and X-ray observables.  We then discuss our plan for applying our method to real data.

\subsection{The impact of projection on multiwavelength observables}

The impact of projection on cluster observables depends on the intrinsic scaling relation, $s \propto M^{\alpha}$.  A larger $\alpha$ corresponds to a stronger signal contrast between high- and low-mass halos, which leads to weaker projection effects.  Here are some estimates of the scaling relations:   
\begin{itemize}
    \item richness $\lambda \propto M^{1.0}$  \cite{murataConstraintsMassRichnessRelation2018, costanziCosmologicalConstraintsY12021}
    \item lensing $\Delta\Sigma \propto M^{0.3-0.7}$ (scale dependent) \citep{wuCosmologyGalaxyCluster2021}
    \item intrinsic SZE $\zeta \propto M^{1.5 - 1.8}$  \cite{costanziCosmologicalConstraintsY12021}
    \item X-ray photon counts $C \propto M^{1.6-2.0}$  \cite{chiuCosmologicalConstraintsGalaxy2023}
\end{itemize}
Given the lower power index of richness and lensing, we expect that the projection effect would be stronger for them than for gas observables.

Physically speaking, richness and lensing projection are dominated by galaxies and matter in line-of-sight filaments [e.g., Fig.~13 in \citet{sunayamaImpactProjectionEffects2020}]. In contrast, SZE projection is dominated by massive halos aligned along the line of sight
\cite{rozoRedMaPPerIIIDetailed2015}. 
The X-ray observables are determined by electron density squared and are less sensitive to projection than SZE signals. In the future, however, as the sensitivities of SZE and X-ray detections improve, their projection effects might need to be considered.

To investigate possible correlated scatter between gas and optical observables, we plan to use hydrodynamic simulations and baryon-pasting models \cite{osatoBaryonPastingAlgorithm2022b}, which assign gas pressure to dark matter halos.  Using these simulations, we plan to integrate the gas pressure along the line of sight to incorporate projection effects into SZE signals.

\subsection{Applying our framework to real data}

Before applying our framework to real data, we need to include several extra systematic effects of cluster lensing.

{\bf The impact of baryonic physics on lensing.}  Our stacked lensing signal is derived from the gravity-only \darkemu.  However, the halo mass density profiles at small scales are modified by baryonic physics \cite{hensonImpactBaryonsMassive2017, debackereHowBaryonsCan2021, giriEmulationBaryonicEffects2021, 2023arXiv230902920G, sunayamaOpticalClusterCosmology2023}. We plan to include the baryonic effect on the lensing profiles in our modeling, following the analytic {\em baryonification} prescription \cite{giriEmulationBaryonicEffects2021}.

{\bf Cluster miscentering.}  In this work, we assume that all clusters are centered at the density peak of the halos.  For clusters identified in DES $\times$ SPT, we 
rely on optical cluster centers because of the relatively large beam size of SPT.  Cluster centers identified by optical imaging sometimes deviate from the density peak.  We plan to incorporate the miscentering fraction quantified using X-ray data \citep{zhangDarkEnergySurvey2019, 2023arXiv231013207K}.

{\bf Extending to X-ray observables.}
This work uses SZE observables as a case study.  With some modifications in the modeling of the ICM observable, our analysis framework is readily applicable to clusters from the extended ROentgen Survey with an Imaging Telescope Array (eROSITA) data.
We expect that the X-ray observables, such as photon count rates, follow a log-normal distribution at a given mass \cite{chiuCosmologicalConstraintsGalaxy2023}, and we can use the full analytic framework presented in Appendix~\ref{sec:lensing_richness_SZE}.

{\bf Individual vs.~stacked lensing.}
In this work, we use stacked lensing signals as our observable to improve the signal-to-noise ratio.  
Several recent studies advocate cluster-by-cluster analyses instead of the stacked approach \citep{grandisCalibrationBiasScatter2021, 2023arXiv231012213B}. 
In the future, with the high source density of the Roman Space Telescope, we might be able to perform cluster-by-cluster lensing analysis in radial bins.

{\bf Monte Carlo vs.~emulator.}
In this work, we use a Monte Carlo approach to facilitate the multi-dimensional integration of non-Gaussian observable--mass distributions. The sampling speed is sufficient for our purpose.  However, in the future, when we vary cosmological parameters, we plan to speed up the modeling by building emulators \citep{heitmannCoyoteUniverseExtended2014, wibkingCosmologyGalaxygalaxyLensing2020}.

Once we perform an optical $\times$ SZE joint analysis, we will be able to constrain the lensing bias due to optical selection and correct for this bias.  We will be able to apply this correction to existing lensing data vectors and conduct cosmological analysis.  This correction is expected to be a significant part of the error budget in optical cluster cosmology analysis.  Another approach is to include the correlated scatter as free parameters and use the optical $\times$ SZE results to put priors on them.  We expect that either approach can effectively address the projection effects in cluster richness and lensing.

\section{Summary and Conclusions}
\label{sec:summary}

The cluster lensing signals of richness-selected samples tend to suffer from a selection bias; namely, projection effects cause a correlated scatter between cluster richness and lensing at a given mass.  As a result, a cluster sample above a richness threshold tends to have a higher mean lensing signal than we would expect from their underlying halo masses.  This effect, if unaccounted for, can lead to biased mass calibration. 
In this work, we present a multiwavelength solution to this selection bias problem. We develop a framework using Monte Carlo methods to model stacked lensing signals in richness and SZE bins, taking into account the correlated scatter between optical observables. 
In parallel, we generate mock cluster catalogs that self-consistently incorporate projection effects for lensing and richness.
We perform likelihood analyses on the mock lensing signals that are binned simultaneously in richness and SZE, and we infer the underlying scaling relation parameters and correlated scatter between lensing and richness.  %
Our lensing-based mock analysis is able to constrain the mean correlated scatter between lensing and richness. 
The posterior distribution of parameters allows us to recover the unbiased stacked lensing signal, which removes the optical selection bias and can be used to derive unbiased lensing mass.

Our result presents an exciting avenue for constraining the optical selection bias of cluster lensing directly from observational data.  Our analysis framework is ready to be applied to the DES$\times$SPT data, and we expect that, with some modifications for the ICM observable--mass relation, the framework can be directly applied to optical$\times$eROSITA data.  We expect that the calibration and the removal of the optical selection bias will lead to unbiased cluster lensing for optically selected samples, which could be a major progress in cosmology with optically identified clusters.

\section*{Acknowledgements}
We thank Erwin Lau, Titus Nyarko Nde, Gladys Kamau, Johnny Esteves, Zhuowen Ben Zhang, for useful inputs for this work. We thank the comments and suggestions from the DES cluster working group.
We thank Boise State and the University of Arizona for hosting two galaxy cluster workshops in 2022 and 2023. 
We acknowledge the use of the lux supercomputer at UC Santa Cruz, funded by NSF MRI grant AST 1828315.
We thank Lehman Garrison for providing the \abacus simulation suites.
ZCH thanks Marco Gatti for providing comments on the manuscript. 
We acknowledge the use of the following open-source software: 
\code{Corrfunc} \citep{sinhaCORRFUNCSuiteBlazing2020}
\code{numpy} \citep{harrisArrayProgrammingNumPy2020}, 
\code{colossus} \citep{diemerCOLOSSUSPythonToolkit2018},
\code{scipy} \citep{virtanenSciPyFundamentalAlgorithms2020}, \code{astropy} \citep{theastropycollaborationAstropyProjectBuilding2018}, \code{jupyter} \citep{grangerJupyterThinkingStorytelling2021}, \code{matplotlib} \citep{hunterMatplotlib2DGraphics2007}.
ZCH and TJ are supported by the DOE award DE-SC0010107. 
HW is supported by the DOE award DE-SC0021916 and the NASA award 15-WFIRST15-0008. 
ANS is supported by the DOE awards DE-SC0009913 and DE-SC0020247. 

\appendix

\section{Analytical treatment of optical selection bias} \label{app:analytical_derivation}

In this section, we extend \citet{evrardModelMultipropertyGalaxy2014} to derive the stacked lensing in richness and SZE bins, assuming an exponential mass function. We also repeat some equations in Sec.~\ref{sec:analytics}. We denote log halo mass by $m$. We assume that the mass function around some pivot mass follows the relation
\begin{equation}
    P(m) \propto n(m) \propto \exp (- \beta m).
\end{equation}

We define the log observables vector as
\begin{equation}
    \ln \boldsymbol{s}=\left(\ln s_1, \ldots, \ln s_n\right). 
\end{equation}
The joint probability distribution is a multivariate Gaussian
\begin{equation}
    \begin{aligned}
& P(\ln \boldsymbol{s} \mid m)=P(\ln \boldsymbol{s} \mid \ln \mu, \mathsf{C}) \\
& =\frac{\exp \left(-\frac{1}{2}(\ln \boldsymbol{s}-\ln \mu)^{\mathrm{T}} \mathsf{C}^{-1}(\ln \boldsymbol{s}-\ln \mathbf{\mu})\right)}{\sqrt{(2 \pi)^n|\mathsf{C}|}},
\end{aligned}
\end{equation}
with the mean vector 
\begin{equation}
    \ln \mu_i(m)=\left\langle\ln s_i \mid m\right\rangle=\alpha_i m+\pi_i,
\end{equation}
and the covariance matrix
\begin{equation}
    \mathsf{C}_{i j}=\left\langle\left(\ln s_i-\ln \mu_i\right)\left(\ln s_j-\ln \mu_j\right)\right\rangle.
\end{equation}
We define the slope vector $\vecalpha=\left\{\alpha_1, \alpha_2,...\right\}$ and the intercept vector $\boldsymbol{\pi} = \left\{\pi_1,\pi_2, ...\right\}$. With the Bayes theorem, $P(m | \vecs)= P(\vecs | m) P(m) / P(\vecs)$, we obtain the mass distribution of halos selected by observables
\begin{align}
    m | \vecs \sim \mathcal{N}(\avg{m|\vecs},\sigmams) ~ ,
\end{align}
in which
\begin{align}
    \sigma_{m \mid \vecs} & =(\boldsymbol{\alpha}^T \mathsf{C}^{-1} \boldsymbol{\alpha})^{-\frac{1}{2}}                                                \\
    \avg{m|\vecs}                  & =\sigma_{m \mid \vecs}^2\left(\boldsymbol{\alpha}^T \mathsf{C}^{-1}(\vecs-\boldsymbol{\pi})-\beta\right) .
\end{align}
The distribution of any set of observables $\vecs_a$ selected by another set of observables $\vecs_b$ is given by
\begin{equation}
P(\vecs_{a}|\vecs_{b}) = \frac{P(\vecs_{a},\vecs_{b})}{P(\vecs_b)}
 = \frac{n(\vecs_a, \vecs_b)}{n(\vecs_b)} .
\label{eq:conditional_probability}
\end{equation}
To obtain the probability density distribution of any given set of observables $P(\vecs)$, we integrate over $P(\vecs|m)$ over the mass function,
\begin{align}
P(\vecs)& = \int dm P(m) P(\vecs | m) \nonumber \\ \nonumber
& = \frac{1}{(2 \pi)^{N / 2} \sqrt{|\mathsf{C}|}} \exp \left\{-\frac{1}{2}(\vecs-\boldsymbol{\pi})^T \mathsf{C}^{-1}(\vecs-\boldsymbol{\pi})\right\} \div \\
& \frac{1}{\sqrt{2 \pi} \sigma_{m \mid \vecs}} \exp \left\{-\frac{1}{2} \frac{\left(\frac{\vecalpha^T \mathsf{C}}{\alphaCalpha} \vecs-\frac{\beta}{\boldsymbol{\alpha}^T \mathsf{C}^{-1} \boldsymbol{\alpha}}\right)^2}{\sigma_{m \mid \vecs}^2}\right\}
\label{eq:space_density}
\end{align}
With the two equations above, we can find, for example, the lensing signal selected simultaneously by richness and SZE signal.

\subsection{Lensing selected by richness} \label{sec:lensing_lambda}
Here we apply Eq.~(\ref{eq:conditional_probability}) and Eq.~(\ref{eq:space_density}) to two log observables: excess surface density $\ln\Delta\Sigma$ and richness $\ln\lambda$. 
Let us assume that $\ln\lambda$ and $\ln\Mwl$ follow a bivariate Gaussian distribution
\begin{equation}
\ln\Mwl, \ln\lambda
\sim \mathcal{N}\bigg(
\big(
\avg{\ln\Mwl|m},\avg{\ln\lambda|m}\big)
, \mathsf{C}\bigg),
\end{equation}
where the mean values are given by
\beqa
\avg{\ln\Mwl|m} &= \alpha_{\Delta\Sigma} m+  \pi_{\Delta\Sigma} 
\\
\avg{\ln\lambda|m} &=
\alpha_{\lambda} m + \pi_{\lambda} ,
\eeqa 
and the covariance matrix is given by 
\begin{equation}
    \mathsf{C} = 
    \begin{pmatrix} 
    \sigma_{\Delta\Sigma}^2 & \rho\sigma_{\Delta\Sigma}\sigma_{\lambda} \\
    \rho\sigma_{\Delta\Sigma}\sigma_{\lambda} & \sigma_{\lambda}^2 \\
\end{pmatrix}  ~.
\end{equation}
We further define
\begin{equation}
m_{\ln s} \equiv \mlambda 
\end{equation}
for any log observable $\ln s$. This quantity corresponds to the log mass obtained by inverting the scaling relation.

The mean log mass of clusters with richness $\lambda$ is given by
\begin{equation}
    \avg{m|\ln \lambda}=m_{\ln \lambda}-\beta \sigma_{m \mid \ln \lambda}^2  ~.
\end{equation}
We note that the last term corresponds to the downward correction due to the shape of the mass function.

The mean $\ln\Mwl$ of this sample is given by 
\beqa 
    \avg{\ln\Mwl|\ln \lambda } &= \pi_{\ln \Delta\Sigma}+\alpha_{\ln \Delta\Sigma}\avg{m|\ln\lambda}  \nonumber\\
    & + \rho \sigma_{\Delta\Sigma} \frac{\left(m_{\ln\lambda}-\avg{m|\ln\lambda}\right)}{\sigma_{m \mid \ln \lambda}}  ~,
    \label{eq:lensinggivenlambda}
\eeqa
The first two terms on the right-hand side of Eq.~(\ref{eq:lensinggivenlambda}) is the mean $\ln\Mwl$ expected from mean halo mass given the
scaling relation, and the last term is the bias caused by the correlation between $\ln\lambda$ and $\ln\Mwl$ at given halo mass.  In order to accurately reconstruct the lensing mass of clusters selected in a richness bin, we need information about the correlation coefficient $\rho$ between $\ln\lambda$ and $\ln\Mwl$ at a given halo mass, as well as the scatter values. 

\subsection{Lensing selected by the richness and thermal Sunyaev--Zeldovich significance}\label{sec:lensing_richness_SZE}

From Eq.~(\ref{eq:conditional_probability}), the lensing signal selected by the richness and SZE signal is given by
\begin{align}
    \label{eq:lensinggivenlamzeta}
\avg{\ln\Mwl | \ln\lambda, \ln\zeta} = &\pi_{\Delta\Sigma} + \alpha_{\Delta\Sigma} \avg{m | \ln\lambda, \ln\zeta} \nonumber \\
&+ \rho \sigma_{\Delta\Sigma} \sigmamu{\ln\lambda}
\Bigg(
\frac{m_{\ln\lambda} - \avg{m|\ln\zeta}}{
    \sigmamu{\ln\zeta}^2 + \sigmamu{\ln\lambda}^2
}
\Bigg),
\end{align}

where the mean mass in the bin selected by $\lambda$ and $\zeta$ is
\begin{equation}
\avg{m | \ln \lambda, \ln\zeta}  
=\frac{
    \sigma_{m|\ln\lambda}^{-2}
    m_{\ln\lambda} +
    \sigma_{m|\ln\zeta}^{-2}
    m_{\ln\zeta} - \beta
}{\sigma_{m|\ln\lambda}^{-2} + \sigma_{m|\ln\zeta}^{-2}
} .
\end{equation}

Note that in equation (\ref{eq:lensinggivenlamzeta}), the first two terms are the result we expect from the scaling relations, and the third term is the bias term caused by the correlation between $\ln\Mwl$ and $\ln\lambda$ at the given mass. 

Note that the bias term is not always positive. It depends on the relation between the mean mass estimate from $\ln\lambda$ and $\ln\zeta$. In the low $\ln\lambda$ and high $\ln\zeta$ bin, this term is likely to be negative, as shown in Fig.~\ref{fig:lensing_abacus}.

To better see the dependence of $\ln\Mwl$ on $\ln\lambda$ and $\ln\zeta$, we reorganize this equation as a bilinear equation in $\ln\lambda$ and $\ln\zeta$. 
\begin{equation}
    \avg{\ln\Mwl |\ln \lambda, \ln \zeta} = A \ln\lambda + B \ln\zeta + C , 
\end{equation}
in which 
\begin{align}
    A & = \frac{\alpha_{\Delta\Sigma} \sigmamu{\ln\zeta}^2 + \rho \sigma_{\Delta\Sigma}\sigmamu{\ln\lambda}}{\alpha_{\lambda}(\sigmamu{\ln\lambda}^2 + \sigmamu{\ln\zeta}^2)}                                                                                      \\
    B & = \frac{\alpha_{\Delta\Sigma} \sigmamu{\ln \lambda}^2 - {\rho} \sigma_{\Delta\Sigma} \sigmamu{\ln\lambda} }{\alpha_{\zeta}(\sigmamu{\ln \lambda}^2 + \sigmamu{\ln\zeta}^2)}                                                                                        \\
    C & = \pi_{\ln\Delta\Sigma} + \frac{\alpha_{\Delta\Sigma}(-\sigmamu{\ln \zeta}^2 \frac{\pi_{\lambda}}{\alpha_{\lambda}}-\sigmamu{\ln \lambda}^2 \frac{\pi_{\zeta}}{\alpha_{\zeta}} - \beta \sigmamu{\ln \lambda}^2\sigmamu{\ln \xi}^2)}{(\sigmamu{\ln\lambda}^2 + \sigmamu{\ln\zeta}^2)}  \nonumber \\
     &+ \frac{\rho \beta \sigma_{\Delta\Sigma}\sigmamu{\ln \lambda}\sigmamu{\ln \zeta}^2}{(\sigmamu{\ln\lambda}^2 + \sigmamu{\ln\zeta}^2)} \nonumber \\
     & - \frac{\rho \sigma_{\Delta\Sigma}\sigmamu{\ln \lambda}\frac{\pi_{\lambda}}{\alpha_{\lambda}}+ \rho \sigma_{\Delta\Sigma}\sigmamu{\ln\lambda}\frac{\pi_{\zeta}}{\alpha_{\zeta}}}{(\sigmamu{\ln\lambda}^2 + \sigmamu{\ln\zeta}^2)}
\end{align}
Since taking the average is a linear operator, the mean lensing signal in a bin is given by 
\begin{equation}
    \braket{\ln\Mwl | \lambda  \text{ bin},\zeta  \text{ bin}} = A\avg{\ln\lambda} + B\avg{\ln\zeta} + C ~,
\end{equation}
in which $\avg{\ln\lambda}$ and $\avg{\ln\zeta}$ are the mean $\ln\lambda$ and mean $\ln\zeta$ in the bin.

\section{Converting scaling relations}\label{sec:scaling_relation}
In this section, we convert the scaling relation from \Costanzi to the convention we use in this work. In \Costanzi, the scaling relation between $\lambda$, $\zeta$, and $M_{500c}$ are:
\begin{align}
\avg{\zeta|m}= & \ln \left(\gamma_f A_{\mathrm{SZE}}\right)+B_{\mathrm{SZE}} \ln \left(\frac{M_{500c}}{3 \times 10^{14} \hiMsun}\right) \nonumber \\ 
& +C_{\mathrm{SZE}} \ln \left(\frac{E(z)}{E(0.6)}\right)
\end{align}
and 
\begin{align}
\avg{\lambda|m}= & \ln \left(A_\lambda\right)+B_\lambda \ln \left(\frac{M_{500c}}{3 \times 10^{14} \hiMsun}\right)\nonumber \\
& +C_\lambda \ln \left(\frac{1+z}{1+0.45}\right)
\end{align}

The conventions we use in this work are:
\begin{align}
\avg{\zeta|m} &= \alpha_{\zeta} m + \pi_{\zeta}\\
\avg{\lambda|m} &= \alpha_{\lambda} m + \pi_{\lambda}
\end{align}

Matching the coefficients, we have 
\begin{align}
\alpha_{\zeta} &= B_{\mathrm{SZE}}\\
\pi_{\zeta} &= \ln(\gamma_f A_{\mathrm{SZE}}) - B_{\mathrm{SZE}}\ln(3\times10^{14}M_\odot h^{-1}) \\
& +C_{\mathrm{SZE}} \ln \left(\frac{E(z)}{E(0.6)}\right) \nonumber\\
\alpha_{\lambda} &= B_{\lambda}\\
\pi_{\lambda} &= \ln(A_\lambda) - B_\lambda\ln(3\times10^{14}M_\odot h^{-1}) +C_{\mathrm{\lambda}} \ln \left(\frac{E(z)}{E(0.6)}\right)
\end{align}

We propagate the errors in $A_{\mathrm{SZE}}, B_{\mathrm{SZE}}, C_{\mathrm{SZE}}, A_{\mathrm{\lambda}}, B_{\mathrm{\lambda}}, C_{\mathrm{\lambda}}$ to the errors in $\alpha_{\lambda}, \pi_{\lambda}, \alpha_{\zeta}, \pi_{\zeta}$ with the simple error propagation formula. Assume $f$ is a function of $x,y,z...$, the total error in $f$ propagated from the errors in the independent variables is
\begin{equation}
    s_f=\sqrt{\left(\frac{\partial f}{\partial x}\right)^2 s_x^2+\left(\frac{\partial f}{\partial y}\right)^2 s_y^2+\left(\frac{\partial f}{\partial z}\right)^2 s_z^2+\cdots}  ~.
\end{equation}
The results are 
\begin{align}
\sigma_{\alpha_\zeta} &= \sigma_{B_{SZE}}\\ 
\sigma_{\pi_\zeta} &= \biggl\{\bigg(\frac{\gamma_f}{\gamma_f A_{SZE}}\bigg)^2 \sigma_{A_{SZE}}^2 + \mpivotcostanzi^2 \sigma_{B_{SZE}}^2 \nonumber \\ 
& + \growthcostanzi^2 \sigma_{C_{SZE}}^2\biggr\}^{\frac{1}{2}}\\
\sigma_{\alpha_\lambda} &= \sigma_{B_\lambda}\\
\sigma_{\pi_\lambda} &= \biggl\{\bigg(\frac{\sigma_{A_{\lambda}}}{A_\lambda}\bigg)^2  + \mpivotcostanzi^2 \sigma_{B_\lambda}^2 \nonumber \\ 
& + \growthcostanzi^2 \sigma_{C_\lambda}^2\biggr\}^{\frac{1}{2}} ~.
\end{align}
We assume the SPT field depth $\gamma_f$ to be 1. The prior we use is the DES-NC + SPT-OMR + PRJ model in Table IV in \Costanzi. 

\section{Maximum likelihood estimator for the fiducial richness--mass relations} \label{sec:mle}

\newcommand{\dd}{{\rm d}}
\newcommand{\Ndetj}{\avg{N_{{\rm det},j}}}
\newcommand{\Ndet}{\avg{N_{{\rm det}}}}

The cluster richness in our mock catalog is calculated using counts-in-cylinders, while our analysis framework assumes a power-law richness--mass relation.  To facilitate the comparison, we derive the fiducial power-law richness--mass relation that describes our mock data.  Since the richness is truncated at $\lambda=20$, we use a maximum likelihood estimator for estimating the scaling relation \cite{mantzObservedGrowthMassive2010, mantzCopingSelectionEffects2019}.

Let us denote the halo mass function (number per log-mass per cubic Mpc) as $\dd n/\dd m$ and the richness--mass relation as $P(\ln\lambda|m)$.
Let us assume infinitesimal log richness bins and log mass bins.  
The expected number of clusters in such an infinitesimal bin $j$ is given by
\beqa
\Ndetj 
& \propto
\begin{dcases}
 P(\ln\lambda_j, m_j
),  \quad \mbox{~if $\lambda_j>\lambda_{\rm th}$ and $m_j >m_{\rm th}$}\\
  0, \quad \mbox{~otherwise}.
\end{dcases}
\eeqa
The normalization is unimportant because we will consider the negative log-likelihood below.  The total number of clusters above the detection threshold is given by
\begin{equation}
\Ndet = \int_{\ln\lambda_{\rm th}}^{\infty} \dd \ln\lambda \int_{m_{\rm th}}^{\infty}
\dd m \left(\frac{\dd N}{\dd m}\right) P(\ln\lambda|m)
\label{eq:Ndet}
\end{equation}
Here, $\dd N/\dd m$ is the halo number per log mass interval (not divided by the volume). 

The negative log-likelihood function, up to a constant, is given by
\begin{equation}
-\ln\mathcal{L}
\big(\{ \ln\lambda_j, m_j \} | \boldsymbol{\Theta} \big) = 
\avg{N_{\rm det} | \boldsymbol{\Theta}} - 
\sum_j %
\ln P(\ln\lambda_j, m_j|\boldsymbol{\Theta}) ,
\label{eq:MLE}
\end{equation}
where $\boldsymbol{\Theta}$ are the scaling relation parameters. 
The last term  includes the contribution from the halo mass function
\begin{equation}
-\ln P(\ln\lambda_j,m_j|\boldsymbol{\Theta})
= -\ln \left(\frac{\dd N}{\dd m}\right) 
- \ln P(\ln\lambda | m) .
\end{equation}
We minimize the negative log-likelihood function given by Eq.~(\ref{eq:MLE})
to derive the scaling parameters $\boldsymbol{\Theta}$.

\section{Scatter of lensing signal at a given halo mass} \label{sec:dsigma_scatter}

For the scatter of $\ln \Delta\Sigma(\rp)$ at a given mass, we measure it with the \textsc{Cardinal} simulation \citep{toBuzzardCardinalImproved2023a}.  The \textsc{Cardinal} simulation provides the noisy $\Delta\Sigma(\rp)$ profile for individual halos measured from dark matter particles, and we use \textsc{DarkEmulator} to measure the mean profile $\avg{\Delta\Sigma(\rp)|m}$.  The fractional scatter of an individual halo in \textsc{Cardinal} can be calculated with 
\begin{equation}
    \sigma_{\Delta\Sigma}(\rp)\approx \frac{\Delta\Sigma(\rp) - \avg{\Delta\Sigma(\rp)|m}}{\avg{\Delta\Sigma(\rp)|m}}.
\end{equation}

Using the median as the estimator, we take the mean of the scatter in the 1-halo and 2-halo regimes, respectively. We use the mean fractional scatters of $\Delta\Sigma$ as first-order estimators of the mean scatter in $\ln \Delta\Sigma$ in the 1-halo and 2-halo regimes. The scatter in $\ln \Delta\Sigma$ values we measure are 0.17 and 0.62 in 1-halo and 2-halo. We have tested that the final constraints on the bias parameters $\omega$ are not sensitive to the scatter in $\ln \Delta\Sigma$. 

\bibliographystyle{apsrev}

\begin{thebibliography}{77}
    \expandafter\ifx\csname natexlab\endcsname\relax\def\natexlab#1{#1}\fi
    \expandafter\ifx\csname bibnamefont\endcsname\relax
      \def\bibnamefont#1{#1}\fi
    \expandafter\ifx\csname bibfnamefont\endcsname\relax
      \def\bibfnamefont#1{#1}\fi
    \expandafter\ifx\csname citenamefont\endcsname\relax
      \def\citenamefont#1{#1}\fi
    \expandafter\ifx\csname url\endcsname\relax
      \def\url#1{\texttt{#1}}\fi
    \expandafter\ifx\csname urlprefix\endcsname\relax\def\urlprefix{URL }\fi
    \providecommand{\bibinfo}[2]{#2}
    \providecommand{\eprint}[2][]{\url{#2}}
    
    \bibitem[{\citenamefont{Allen et~al.}(2011)\citenamefont{Allen, Evrard, and Mantz}}]{allenCosmologicalParametersObservations2011}
    \bibinfo{author}{\bibfnamefont{S.~W.} \bibnamefont{Allen}}, \bibinfo{author}{\bibfnamefont{A.~E.} \bibnamefont{Evrard}}, \bibnamefont{and} \bibinfo{author}{\bibfnamefont{A.~B.} \bibnamefont{Mantz}}, \bibinfo{journal}{Annual Review of Astronomy and Astrophysics} \textbf{\bibinfo{volume}{49}}, \bibinfo{pages}{409} (\bibinfo{year}{2011}), ISSN \bibinfo{issn}{0066-4146, 1545-4282}, \eprint{1103.4829}.
    
    \bibitem[{\citenamefont{Weinberg et~al.}(2013)\citenamefont{Weinberg, Mortonson, Eisenstein, Hirata, Riess, and Rozo}}]{weinbergObservationalProbesCosmic2013}
    \bibinfo{author}{\bibfnamefont{D.~H.} \bibnamefont{Weinberg}}, \bibinfo{author}{\bibfnamefont{M.~J.} \bibnamefont{Mortonson}}, \bibinfo{author}{\bibfnamefont{D.~J.} \bibnamefont{Eisenstein}}, \bibinfo{author}{\bibfnamefont{C.}~\bibnamefont{Hirata}}, \bibinfo{author}{\bibfnamefont{A.~G.} \bibnamefont{Riess}}, \bibnamefont{and} \bibinfo{author}{\bibfnamefont{E.}~\bibnamefont{Rozo}}, \bibinfo{journal}{Physics Reports} \textbf{\bibinfo{volume}{530}}, \bibinfo{pages}{87} (\bibinfo{year}{2013}), ISSN \bibinfo{issn}{0370-1573}.
    
    \bibitem[{\citenamefont{Huterer et~al.}(2015)\citenamefont{Huterer, Kirkby, Bean, Connolly, Dawson, Dodelson, Evrard, Jain, Jarvis, Linder et~al.}}]{hutererGrowthCosmicStructure2015}
    \bibinfo{author}{\bibfnamefont{D.}~\bibnamefont{Huterer}}, \bibinfo{author}{\bibfnamefont{D.}~\bibnamefont{Kirkby}}, \bibinfo{author}{\bibfnamefont{R.}~\bibnamefont{Bean}}, \bibinfo{author}{\bibfnamefont{A.}~\bibnamefont{Connolly}}, \bibinfo{author}{\bibfnamefont{K.}~\bibnamefont{Dawson}}, \bibinfo{author}{\bibfnamefont{S.}~\bibnamefont{Dodelson}}, \bibinfo{author}{\bibfnamefont{A.}~\bibnamefont{Evrard}}, \bibinfo{author}{\bibfnamefont{B.}~\bibnamefont{Jain}}, \bibinfo{author}{\bibfnamefont{M.}~\bibnamefont{Jarvis}}, \bibinfo{author}{\bibfnamefont{E.}~\bibnamefont{Linder}}, \bibnamefont{et~al.}, \bibinfo{journal}{Astroparticle Physics} \textbf{\bibinfo{volume}{63}}, \bibinfo{pages}{23} (\bibinfo{year}{2015}), ISSN \bibinfo{issn}{0927-6505}.
    
    \bibitem[{\citenamefont{Rykoff et~al.}(2014)\citenamefont{Rykoff, Rozo, Busha, Cunha, Finoguenov, Evrard, Hao, Koester, Leauthaud, Nord et~al.}}]{rykoffRedMaPPerAlgorithmSDSS2014}
    \bibinfo{author}{\bibfnamefont{E.~S.} \bibnamefont{Rykoff}}, \bibinfo{author}{\bibfnamefont{E.}~\bibnamefont{Rozo}}, \bibinfo{author}{\bibfnamefont{M.~T.} \bibnamefont{Busha}}, \bibinfo{author}{\bibfnamefont{C.~E.} \bibnamefont{Cunha}}, \bibinfo{author}{\bibfnamefont{A.}~\bibnamefont{Finoguenov}}, \bibinfo{author}{\bibfnamefont{A.}~\bibnamefont{Evrard}}, \bibinfo{author}{\bibfnamefont{J.}~\bibnamefont{Hao}}, \bibinfo{author}{\bibfnamefont{B.~P.} \bibnamefont{Koester}}, \bibinfo{author}{\bibfnamefont{A.}~\bibnamefont{Leauthaud}}, \bibinfo{author}{\bibfnamefont{B.}~\bibnamefont{Nord}}, \bibnamefont{et~al.}, \bibinfo{journal}{The Astrophysical Journal} \textbf{\bibinfo{volume}{785}}, \bibinfo{pages}{104} (\bibinfo{year}{2014}), ISSN \bibinfo{issn}{0004-637X, 1538-4357}, \eprint{1303.3562}.
    
    \bibitem[{\citenamefont{Mantz et~al.}(2010)\citenamefont{Mantz, Allen, Rapetti, and Ebeling}}]{mantzObservedGrowthMassive2010}
    \bibinfo{author}{\bibfnamefont{A.}~\bibnamefont{Mantz}}, \bibinfo{author}{\bibfnamefont{S.~W.} \bibnamefont{Allen}}, \bibinfo{author}{\bibfnamefont{D.}~\bibnamefont{Rapetti}}, \bibnamefont{and} \bibinfo{author}{\bibfnamefont{H.}~\bibnamefont{Ebeling}}, \bibinfo{journal}{Monthly Notices of the Royal Astronomical Society} \textbf{\bibinfo{volume}{406}}, \bibinfo{pages}{1759} (\bibinfo{year}{2010}), ISSN \bibinfo{issn}{0035-8711}.
    
    \bibitem[{\citenamefont{Chiu et~al.}(2023)\citenamefont{Chiu, Klein, Mohr, and Bocquet}}]{chiuCosmologicalConstraintsGalaxy2023}
    \bibinfo{author}{\bibfnamefont{I.~N.} \bibnamefont{Chiu}}, \bibinfo{author}{\bibfnamefont{M.}~\bibnamefont{Klein}}, \bibinfo{author}{\bibfnamefont{J.}~\bibnamefont{Mohr}}, \bibnamefont{and} \bibinfo{author}{\bibfnamefont{S.}~\bibnamefont{Bocquet}}, \bibinfo{journal}{Monthly Notices of the Royal Astronomical Society} \textbf{\bibinfo{volume}{522}}, \bibinfo{pages}{1601} (\bibinfo{year}{2023}), ISSN \bibinfo{issn}{0035-8711}.
    
    \bibitem[{\citenamefont{Sunyaev and Zeldovich}(1970)}]{sunyaevSmallscaleFluctuationsRelic1970}
    \bibinfo{author}{\bibfnamefont{R.~A.} \bibnamefont{Sunyaev}} \bibnamefont{and} \bibinfo{author}{\bibfnamefont{{\relax Ya}.~B.} \bibnamefont{Zeldovich}}, \bibinfo{journal}{Astrophysics and Space Science} \textbf{\bibinfo{volume}{7}}, \bibinfo{pages}{3} (\bibinfo{year}{1970}), ISSN \bibinfo{issn}{1572-946X}.
    
    \bibitem[{\citenamefont{Saro et~al.}(2015)\citenamefont{Saro, Bocquet, Rozo, Benson, Mohr, Rykoff, {Soares-Santos}, Bleem, Dodelson, Melchior et~al.}}]{saroConstraintsRichnessMass2015}
    \bibinfo{author}{\bibfnamefont{A.}~\bibnamefont{Saro}}, \bibinfo{author}{\bibfnamefont{S.}~\bibnamefont{Bocquet}}, \bibinfo{author}{\bibfnamefont{E.}~\bibnamefont{Rozo}}, \bibinfo{author}{\bibfnamefont{B.~A.} \bibnamefont{Benson}}, \bibinfo{author}{\bibfnamefont{J.}~\bibnamefont{Mohr}}, \bibinfo{author}{\bibfnamefont{E.~S.} \bibnamefont{Rykoff}}, \bibinfo{author}{\bibfnamefont{M.}~\bibnamefont{{Soares-Santos}}}, \bibinfo{author}{\bibfnamefont{L.}~\bibnamefont{Bleem}}, \bibinfo{author}{\bibfnamefont{S.}~\bibnamefont{Dodelson}}, \bibinfo{author}{\bibfnamefont{P.}~\bibnamefont{Melchior}}, \bibnamefont{et~al.}, \bibinfo{journal}{Monthly Notices of the Royal Astronomical Society} \textbf{\bibinfo{volume}{454}}, \bibinfo{pages}{2305} (\bibinfo{year}{2015}), ISSN \bibinfo{issn}{0035-8711}.
    
    \bibitem[{\citenamefont{Saro et~al.}(2017)\citenamefont{Saro, Bocquet, Mohr, Rozo, Benson, Dodelson, Rykoff, Bleem, Abbott, Abdalla et~al.}}]{saroOpticalSZEScalingRelations2017}
    \bibinfo{author}{\bibfnamefont{A.}~\bibnamefont{Saro}}, \bibinfo{author}{\bibfnamefont{S.}~\bibnamefont{Bocquet}}, \bibinfo{author}{\bibfnamefont{J.}~\bibnamefont{Mohr}}, \bibinfo{author}{\bibfnamefont{E.}~\bibnamefont{Rozo}}, \bibinfo{author}{\bibfnamefont{B.~A.} \bibnamefont{Benson}}, \bibinfo{author}{\bibfnamefont{S.}~\bibnamefont{Dodelson}}, \bibinfo{author}{\bibfnamefont{E.~S.} \bibnamefont{Rykoff}}, \bibinfo{author}{\bibfnamefont{L.}~\bibnamefont{Bleem}}, \bibinfo{author}{\bibfnamefont{T.~M.~C.} \bibnamefont{Abbott}}, \bibinfo{author}{\bibfnamefont{F.~B.} \bibnamefont{Abdalla}}, \bibnamefont{et~al.}, \bibinfo{journal}{Monthly Notices of the Royal Astronomical Society} \textbf{\bibinfo{volume}{468}}, \bibinfo{pages}{3347} (\bibinfo{year}{2017}), ISSN \bibinfo{issn}{0035-8711, 1365-2966}, \eprint{1605.08770}.
    
    \bibitem[{\citenamefont{Bocquet et~al.}(2019)\citenamefont{Bocquet, Dietrich, Schrabback, Bleem, Klein, Allen, Applegate, Ashby, Bautz, Bayliss et~al.}}]{bocquetClusterCosmologyConstraints2019}
    \bibinfo{author}{\bibfnamefont{S.}~\bibnamefont{Bocquet}}, \bibinfo{author}{\bibfnamefont{J.~P.} \bibnamefont{Dietrich}}, \bibinfo{author}{\bibfnamefont{T.}~\bibnamefont{Schrabback}}, \bibinfo{author}{\bibfnamefont{L.~E.} \bibnamefont{Bleem}}, \bibinfo{author}{\bibfnamefont{M.}~\bibnamefont{Klein}}, \bibinfo{author}{\bibfnamefont{S.~W.} \bibnamefont{Allen}}, \bibinfo{author}{\bibfnamefont{D.~E.} \bibnamefont{Applegate}}, \bibinfo{author}{\bibfnamefont{M.~L.~N.} \bibnamefont{Ashby}}, \bibinfo{author}{\bibfnamefont{M.}~\bibnamefont{Bautz}}, \bibinfo{author}{\bibfnamefont{M.}~\bibnamefont{Bayliss}}, \bibnamefont{et~al.}, \bibinfo{journal}{The Astrophysical Journal} \textbf{\bibinfo{volume}{878}}, \bibinfo{pages}{55} (\bibinfo{year}{2019}), ISSN \bibinfo{issn}{1538-4357}, \eprint{1812.01679}.
    
    \bibitem[{\citenamefont{Costanzi et~al.}(2021)\citenamefont{Costanzi, Saro, Bocquet, Abbott, Aguena, Allam, Amara, Annis, Avila, Bacon et~al.}}]{costanziCosmologicalConstraintsY12021}
    \bibinfo{author}{\bibfnamefont{M.}~\bibnamefont{Costanzi}}, \bibinfo{author}{\bibfnamefont{A.}~\bibnamefont{Saro}}, \bibinfo{author}{\bibfnamefont{S.}~\bibnamefont{Bocquet}}, \bibinfo{author}{\bibfnamefont{T.~M.~C.} \bibnamefont{Abbott}}, \bibinfo{author}{\bibfnamefont{M.}~\bibnamefont{Aguena}}, \bibinfo{author}{\bibfnamefont{S.}~\bibnamefont{Allam}}, \bibinfo{author}{\bibfnamefont{A.}~\bibnamefont{Amara}}, \bibinfo{author}{\bibfnamefont{J.}~\bibnamefont{Annis}}, \bibinfo{author}{\bibfnamefont{S.}~\bibnamefont{Avila}}, \bibinfo{author}{\bibfnamefont{D.}~\bibnamefont{Bacon}}, \bibnamefont{et~al.}, \bibinfo{journal}{Physical Review D} \textbf{\bibinfo{volume}{103}}, \bibinfo{pages}{043522} (\bibinfo{year}{2021}), ISSN \bibinfo{issn}{2470-0010, 2470-0029}, \eprint{2010.13800}.
    
    \bibitem[{\citenamefont{Pratt et~al.}(2019)\citenamefont{Pratt, Arnaud, Biviano, Eckert, Ettori, Nagai, Okabe, and Reiprich}}]{prattGalaxyClusterMass2019}
    \bibinfo{author}{\bibfnamefont{G.~W.} \bibnamefont{Pratt}}, \bibinfo{author}{\bibfnamefont{M.}~\bibnamefont{Arnaud}}, \bibinfo{author}{\bibfnamefont{A.}~\bibnamefont{Biviano}}, \bibinfo{author}{\bibfnamefont{D.}~\bibnamefont{Eckert}}, \bibinfo{author}{\bibfnamefont{S.}~\bibnamefont{Ettori}}, \bibinfo{author}{\bibfnamefont{D.}~\bibnamefont{Nagai}}, \bibinfo{author}{\bibfnamefont{N.}~\bibnamefont{Okabe}}, \bibnamefont{and} \bibinfo{author}{\bibfnamefont{T.~H.} \bibnamefont{Reiprich}}, \bibinfo{journal}{Space Science Reviews} \textbf{\bibinfo{volume}{215}}, \bibinfo{pages}{25} (\bibinfo{year}{2019}), ISSN \bibinfo{issn}{0038-6308}.
    
    \bibitem[{\citenamefont{Umetsu}(2020)}]{umetsuClustergalaxyWeakLensing2020}
    \bibinfo{author}{\bibfnamefont{K.}~\bibnamefont{Umetsu}}, \bibinfo{journal}{The Astronomy and Astrophysics Review} \textbf{\bibinfo{volume}{28}}, \bibinfo{pages}{7} (\bibinfo{year}{2020}), ISSN \bibinfo{issn}{0935-4956, 1432-0754}, \eprint{2007.00506}.
    
    \bibitem[{\citenamefont{Becker and Kravtsov}(2011)}]{beckerACCURACYWEAKLENSINGCLUSTER2011}
    \bibinfo{author}{\bibfnamefont{M.~R.} \bibnamefont{Becker}} \bibnamefont{and} \bibinfo{author}{\bibfnamefont{A.~V.} \bibnamefont{Kravtsov}}, \bibinfo{journal}{The Astrophysical Journal} \textbf{\bibinfo{volume}{740}}, \bibinfo{pages}{25} (\bibinfo{year}{2011}), ISSN \bibinfo{issn}{0004-637X}.
    
    \bibitem[{\citenamefont{Wu et~al.}(2019)\citenamefont{Wu, Weinberg, Salcedo, Wibking, and Zu}}]{wuCovarianceMatricesGalaxy2019}
    \bibinfo{author}{\bibfnamefont{H.-Y.} \bibnamefont{Wu}}, \bibinfo{author}{\bibfnamefont{D.~H.} \bibnamefont{Weinberg}}, \bibinfo{author}{\bibfnamefont{A.~N.} \bibnamefont{Salcedo}}, \bibinfo{author}{\bibfnamefont{B.~D.} \bibnamefont{Wibking}}, \bibnamefont{and} \bibinfo{author}{\bibfnamefont{Y.}~\bibnamefont{Zu}}, \bibinfo{journal}{Monthly Notices of the Royal Astronomical Society} \textbf{\bibinfo{volume}{490}}, \bibinfo{pages}{2606} (\bibinfo{year}{2019}), ISSN \bibinfo{issn}{0035-8711, 1365-2966}, \eprint{1907.06611}.
    
    \bibitem[{\citenamefont{Johnston et~al.}(2007)\citenamefont{Johnston, Sheldon, Wechsler, Rozo, Koester, Frieman, McKay, Evrard, Becker, and Annis}}]{johnstonCrosscorrelationWeakLensing2007}
    \bibinfo{author}{\bibfnamefont{D.~E.} \bibnamefont{Johnston}}, \bibinfo{author}{\bibfnamefont{E.~S.} \bibnamefont{Sheldon}}, \bibinfo{author}{\bibfnamefont{R.~H.} \bibnamefont{Wechsler}}, \bibinfo{author}{\bibfnamefont{E.}~\bibnamefont{Rozo}}, \bibinfo{author}{\bibfnamefont{B.~P.} \bibnamefont{Koester}}, \bibinfo{author}{\bibfnamefont{J.~A.} \bibnamefont{Frieman}}, \bibinfo{author}{\bibfnamefont{T.~A.} \bibnamefont{McKay}}, \bibinfo{author}{\bibfnamefont{A.~E.} \bibnamefont{Evrard}}, \bibinfo{author}{\bibfnamefont{M.~R.} \bibnamefont{Becker}}, \bibnamefont{and} \bibinfo{author}{\bibfnamefont{J.}~\bibnamefont{Annis}} (\bibinfo{year}{2007}), \eprint{0709.1159}.
    
    \bibitem[{\citenamefont{Simet et~al.}(2017)\citenamefont{Simet, McClintock, Mandelbaum, Rozo, Rykoff, Sheldon, and Wechsler}}]{simetWeakLensingMeasurement2017}
    \bibinfo{author}{\bibfnamefont{M.}~\bibnamefont{Simet}}, \bibinfo{author}{\bibfnamefont{T.}~\bibnamefont{McClintock}}, \bibinfo{author}{\bibfnamefont{R.}~\bibnamefont{Mandelbaum}}, \bibinfo{author}{\bibfnamefont{E.}~\bibnamefont{Rozo}}, \bibinfo{author}{\bibfnamefont{E.}~\bibnamefont{Rykoff}}, \bibinfo{author}{\bibfnamefont{E.}~\bibnamefont{Sheldon}}, \bibnamefont{and} \bibinfo{author}{\bibfnamefont{R.~H.} \bibnamefont{Wechsler}}, \bibinfo{journal}{Monthly Notices of the Royal Astronomical Society} \textbf{\bibinfo{volume}{466}}, \bibinfo{pages}{3103} (\bibinfo{year}{2017}), ISSN \bibinfo{issn}{0035-8711}.
    
    \bibitem[{\citenamefont{Melchior et~al.}(2017)\citenamefont{Melchior, Gruen, McClintock, Varga, Sheldon, Rozo, Amara, Becker, Benson, Bermeo et~al.}}]{melchiorWeaklensingMassCalibration2017}
    \bibinfo{author}{\bibfnamefont{P.}~\bibnamefont{Melchior}}, \bibinfo{author}{\bibfnamefont{D.}~\bibnamefont{Gruen}}, \bibinfo{author}{\bibfnamefont{T.}~\bibnamefont{McClintock}}, \bibinfo{author}{\bibfnamefont{T.~N.} \bibnamefont{Varga}}, \bibinfo{author}{\bibfnamefont{E.}~\bibnamefont{Sheldon}}, \bibinfo{author}{\bibfnamefont{E.}~\bibnamefont{Rozo}}, \bibinfo{author}{\bibfnamefont{A.}~\bibnamefont{Amara}}, \bibinfo{author}{\bibfnamefont{M.~R.} \bibnamefont{Becker}}, \bibinfo{author}{\bibfnamefont{B.~A.} \bibnamefont{Benson}}, \bibinfo{author}{\bibfnamefont{A.}~\bibnamefont{Bermeo}}, \bibnamefont{et~al.}, \bibinfo{journal}{Monthly Notices of the Royal Astronomical Society} \textbf{\bibinfo{volume}{469}}, \bibinfo{pages}{4899} (\bibinfo{year}{2017}), ISSN \bibinfo{issn}{0035-8711}.
    
    \bibitem[{\citenamefont{Murata et~al.}(2018)\citenamefont{Murata, Nishimichi, Takada, Miyatake, Shirasaki, More, Takahashi, and Osato}}]{murataConstraintsMassRichnessRelation2018}
    \bibinfo{author}{\bibfnamefont{R.}~\bibnamefont{Murata}}, \bibinfo{author}{\bibfnamefont{T.}~\bibnamefont{Nishimichi}}, \bibinfo{author}{\bibfnamefont{M.}~\bibnamefont{Takada}}, \bibinfo{author}{\bibfnamefont{H.}~\bibnamefont{Miyatake}}, \bibinfo{author}{\bibfnamefont{M.}~\bibnamefont{Shirasaki}}, \bibinfo{author}{\bibfnamefont{S.}~\bibnamefont{More}}, \bibinfo{author}{\bibfnamefont{R.}~\bibnamefont{Takahashi}}, \bibnamefont{and} \bibinfo{author}{\bibfnamefont{K.}~\bibnamefont{Osato}}, \bibinfo{journal}{The Astrophysical Journal} \textbf{\bibinfo{volume}{854}}, \bibinfo{pages}{120} (\bibinfo{year}{2018}), ISSN \bibinfo{issn}{0004-637X}.
    
    \bibitem[{\citenamefont{McClintock et~al.}(2019)\citenamefont{McClintock, Varga, Gruen, Rozo, Rykoff, Shin, Melchior, DeRose, Seitz, Dietrich et~al.}}]{mcclintockDarkEnergySurvey2019}
    \bibinfo{author}{\bibfnamefont{T.}~\bibnamefont{McClintock}}, \bibinfo{author}{\bibfnamefont{T.~N.} \bibnamefont{Varga}}, \bibinfo{author}{\bibfnamefont{D.}~\bibnamefont{Gruen}}, \bibinfo{author}{\bibfnamefont{E.}~\bibnamefont{Rozo}}, \bibinfo{author}{\bibfnamefont{E.~S.} \bibnamefont{Rykoff}}, \bibinfo{author}{\bibfnamefont{T.}~\bibnamefont{Shin}}, \bibinfo{author}{\bibfnamefont{P.}~\bibnamefont{Melchior}}, \bibinfo{author}{\bibfnamefont{J.}~\bibnamefont{DeRose}}, \bibinfo{author}{\bibfnamefont{S.}~\bibnamefont{Seitz}}, \bibinfo{author}{\bibfnamefont{J.~P.} \bibnamefont{Dietrich}}, \bibnamefont{et~al.}, \bibinfo{journal}{Monthly Notices of the Royal Astronomical Society} \textbf{\bibinfo{volume}{482}}, \bibinfo{pages}{1352} (\bibinfo{year}{2019}), ISSN \bibinfo{issn}{0035-8711, 1365-2966}, \eprint{1805.00039}.
    
    \bibitem[{\citenamefont{{Bocquet} et~al.}(2023)\citenamefont{{Bocquet}, {Grandis}, {Bleem}, {Klein}, {Mohr}, {Aguena}, {Alarcon}, {Allam}, {Allen}, {Alves} et~al.}}]{2023arXiv231012213B}
    \bibinfo{author}{\bibfnamefont{S.}~\bibnamefont{{Bocquet}}}, \bibinfo{author}{\bibfnamefont{S.}~\bibnamefont{{Grandis}}}, \bibinfo{author}{\bibfnamefont{L.~E.} \bibnamefont{{Bleem}}}, \bibinfo{author}{\bibfnamefont{M.}~\bibnamefont{{Klein}}}, \bibinfo{author}{\bibfnamefont{J.~J.} \bibnamefont{{Mohr}}}, \bibinfo{author}{\bibfnamefont{M.}~\bibnamefont{{Aguena}}}, \bibinfo{author}{\bibfnamefont{A.}~\bibnamefont{{Alarcon}}}, \bibinfo{author}{\bibfnamefont{S.}~\bibnamefont{{Allam}}}, \bibinfo{author}{\bibfnamefont{S.~W.} \bibnamefont{{Allen}}}, \bibinfo{author}{\bibfnamefont{O.}~\bibnamefont{{Alves}}}, \bibnamefont{et~al.}, \bibinfo{journal}{arXiv e-prints} \bibinfo{eid}{arXiv:2310.12213} (\bibinfo{year}{2023}), \eprint{2310.12213}.
    
    \bibitem[{\citenamefont{Cohn et~al.}(2007)\citenamefont{Cohn, Evrard, White, Croton, and Ellingson}}]{cohnRedsequenceClusterFinding2007}
    \bibinfo{author}{\bibfnamefont{J.~D.} \bibnamefont{Cohn}}, \bibinfo{author}{\bibfnamefont{A.~E.} \bibnamefont{Evrard}}, \bibinfo{author}{\bibfnamefont{M.}~\bibnamefont{White}}, \bibinfo{author}{\bibfnamefont{D.}~\bibnamefont{Croton}}, \bibnamefont{and} \bibinfo{author}{\bibfnamefont{E.}~\bibnamefont{Ellingson}}, \bibinfo{journal}{Monthly Notices of the Royal Astronomical Society} \textbf{\bibinfo{volume}{382}}, \bibinfo{pages}{1738} (\bibinfo{year}{2007}), ISSN \bibinfo{issn}{0035-8711}.
    
    \bibitem[{\citenamefont{Noh and Cohn}(2011)}]{nohGeometryFilamentaryEnvironment2011}
    \bibinfo{author}{\bibfnamefont{Y.}~\bibnamefont{Noh}} \bibnamefont{and} \bibinfo{author}{\bibfnamefont{J.~D.} \bibnamefont{Cohn}}, \bibinfo{journal}{Monthly Notices of the Royal Astronomical Society} \textbf{\bibinfo{volume}{413}}, \bibinfo{pages}{301} (\bibinfo{year}{2011}), ISSN \bibinfo{issn}{0035-8711}.
    
    \bibitem[{\citenamefont{Busch and White}(2017)}]{buschAssemblyBiasSplashback2017}
    \bibinfo{author}{\bibfnamefont{P.}~\bibnamefont{Busch}} \bibnamefont{and} \bibinfo{author}{\bibfnamefont{S.~D.~M.} \bibnamefont{White}}, \bibinfo{journal}{Monthly Notices of the Royal Astronomical Society} \textbf{\bibinfo{volume}{470}}, \bibinfo{pages}{4767} (\bibinfo{year}{2017}), ISSN \bibinfo{issn}{0035-8711, 1365-2966}, \eprint{1702.01682}.
    
    \bibitem[{\citenamefont{Zu et~al.}(2017)\citenamefont{Zu, Mandelbaum, Simet, Rozo, and Rykoff}}]{zuLevelClusterAssembly2017}
    \bibinfo{author}{\bibfnamefont{Y.}~\bibnamefont{Zu}}, \bibinfo{author}{\bibfnamefont{R.}~\bibnamefont{Mandelbaum}}, \bibinfo{author}{\bibfnamefont{M.}~\bibnamefont{Simet}}, \bibinfo{author}{\bibfnamefont{E.}~\bibnamefont{Rozo}}, \bibnamefont{and} \bibinfo{author}{\bibfnamefont{E.~S.} \bibnamefont{Rykoff}}, \bibinfo{journal}{Monthly Notices of the Royal Astronomical Society} \textbf{\bibinfo{volume}{470}}, \bibinfo{pages}{551} (\bibinfo{year}{2017}), ISSN \bibinfo{issn}{0035-8711, 1365-2966}, \eprint{1611.00366}.
    
    \bibitem[{\citenamefont{Sunayama et~al.}(2020)\citenamefont{Sunayama, Park, Takada, Kobayashi, Nishimichi, Kurita, More, Oguri, and Osato}}]{sunayamaImpactProjectionEffects2020}
    \bibinfo{author}{\bibfnamefont{T.}~\bibnamefont{Sunayama}}, \bibinfo{author}{\bibfnamefont{Y.}~\bibnamefont{Park}}, \bibinfo{author}{\bibfnamefont{M.}~\bibnamefont{Takada}}, \bibinfo{author}{\bibfnamefont{Y.}~\bibnamefont{Kobayashi}}, \bibinfo{author}{\bibfnamefont{T.}~\bibnamefont{Nishimichi}}, \bibinfo{author}{\bibfnamefont{T.}~\bibnamefont{Kurita}}, \bibinfo{author}{\bibfnamefont{S.}~\bibnamefont{More}}, \bibinfo{author}{\bibfnamefont{M.}~\bibnamefont{Oguri}}, \bibnamefont{and} \bibinfo{author}{\bibfnamefont{K.}~\bibnamefont{Osato}}, \bibinfo{journal}{Monthly Notices of the Royal Astronomical Society} \textbf{\bibinfo{volume}{496}}, \bibinfo{pages}{4468} (\bibinfo{year}{2020}), ISSN \bibinfo{issn}{0035-8711, 1365-2966}, \eprint{2002.03867}.
    
    \bibitem[{\citenamefont{Wu et~al.}(2022)\citenamefont{Wu, Costanzi, To, Salcedo, Weinberg, Annis, Bocquet, {da~Silva~Pereira}, DeRose, Esteves et~al.}}]{wuOpticalSelectionBias2022}
    \bibinfo{author}{\bibfnamefont{H.-Y.} \bibnamefont{Wu}}, \bibinfo{author}{\bibfnamefont{M.}~\bibnamefont{Costanzi}}, \bibinfo{author}{\bibfnamefont{C.-H.} \bibnamefont{To}}, \bibinfo{author}{\bibfnamefont{A.~N.} \bibnamefont{Salcedo}}, \bibinfo{author}{\bibfnamefont{D.~H.} \bibnamefont{Weinberg}}, \bibinfo{author}{\bibfnamefont{J.}~\bibnamefont{Annis}}, \bibinfo{author}{\bibfnamefont{S.}~\bibnamefont{Bocquet}}, \bibinfo{author}{\bibfnamefont{M.~E.} \bibnamefont{{da~Silva~Pereira}}}, \bibinfo{author}{\bibfnamefont{J.}~\bibnamefont{DeRose}}, \bibinfo{author}{\bibfnamefont{J.}~\bibnamefont{Esteves}}, \bibnamefont{et~al.}, \bibinfo{journal}{Monthly Notices of the Royal Astronomical Society} \textbf{\bibinfo{volume}{515}}, \bibinfo{pages}{4471} (\bibinfo{year}{2022}), ISSN \bibinfo{issn}{0035-8711}.
    
    \bibitem[{\citenamefont{Zhang and Annis}(2022)}]{zhangEffectSelectionTale2022}
    \bibinfo{author}{\bibfnamefont{Y.}~\bibnamefont{Zhang}} \bibnamefont{and} \bibinfo{author}{\bibfnamefont{J.}~\bibnamefont{Annis}}, \bibinfo{journal}{Monthly Notices of the Royal Astronomical Society} \textbf{\bibinfo{volume}{511}}, \bibinfo{pages}{L30} (\bibinfo{year}{2022}), ISSN \bibinfo{issn}{0035-8711}.
    
    \bibitem[{\citenamefont{Zhang et~al.}(2023)\citenamefont{Zhang, Wu, Zhang, Frieman, To, DeRose, Costanzi, Wechsler, Adhikari, Rykoff et~al.}}]{zhangModellingGalaxyCluster2023}
    \bibinfo{author}{\bibfnamefont{Z.}~\bibnamefont{Zhang}}, \bibinfo{author}{\bibfnamefont{H.-Y.} \bibnamefont{Wu}}, \bibinfo{author}{\bibfnamefont{Y.}~\bibnamefont{Zhang}}, \bibinfo{author}{\bibfnamefont{J.}~\bibnamefont{Frieman}}, \bibinfo{author}{\bibfnamefont{C.-H.} \bibnamefont{To}}, \bibinfo{author}{\bibfnamefont{J.}~\bibnamefont{DeRose}}, \bibinfo{author}{\bibfnamefont{M.}~\bibnamefont{Costanzi}}, \bibinfo{author}{\bibfnamefont{R.~H.} \bibnamefont{Wechsler}}, \bibinfo{author}{\bibfnamefont{S.}~\bibnamefont{Adhikari}}, \bibinfo{author}{\bibfnamefont{E.}~\bibnamefont{Rykoff}}, \bibnamefont{et~al.}, \bibinfo{journal}{Monthly Notices of the Royal Astronomical Society} \textbf{\bibinfo{volume}{523}}, \bibinfo{pages}{1994} (\bibinfo{year}{2023}), ISSN \bibinfo{issn}{0035-8711}.
    
    \bibitem[{\citenamefont{Abbott et~al.}(2020)\citenamefont{Abbott, Aguena, Alarcon, Allam, Allen, Annis, Avila, Bacon, Bechtol, Bermeo et~al.}}]{abbottDarkEnergySurvey2020}
    \bibinfo{author}{\bibfnamefont{T.~M.~C.} \bibnamefont{Abbott}}, \bibinfo{author}{\bibfnamefont{M.}~\bibnamefont{Aguena}}, \bibinfo{author}{\bibfnamefont{A.}~\bibnamefont{Alarcon}}, \bibinfo{author}{\bibfnamefont{S.}~\bibnamefont{Allam}}, \bibinfo{author}{\bibfnamefont{S.}~\bibnamefont{Allen}}, \bibinfo{author}{\bibfnamefont{J.}~\bibnamefont{Annis}}, \bibinfo{author}{\bibfnamefont{S.}~\bibnamefont{Avila}}, \bibinfo{author}{\bibfnamefont{D.}~\bibnamefont{Bacon}}, \bibinfo{author}{\bibfnamefont{K.}~\bibnamefont{Bechtol}}, \bibinfo{author}{\bibfnamefont{A.}~\bibnamefont{Bermeo}}, \bibnamefont{et~al.}, \bibinfo{journal}{Physical Review D} \textbf{\bibinfo{volume}{102}}, \bibinfo{pages}{023509} (\bibinfo{year}{2020}), ISSN \bibinfo{issn}{2470-0010, 2470-0029}.
    
    \bibitem[{\citenamefont{Sunayama et~al.}(2023)\citenamefont{Sunayama, Miyatake, Sugiyama, More, Li, Dalal, Rau, Shi, Chiu, Shirasaki et~al.}}]{sunayamaOpticalClusterCosmology2023}
    \bibinfo{author}{\bibfnamefont{T.}~\bibnamefont{Sunayama}}, \bibinfo{author}{\bibfnamefont{H.}~\bibnamefont{Miyatake}}, \bibinfo{author}{\bibfnamefont{S.}~\bibnamefont{Sugiyama}}, \bibinfo{author}{\bibfnamefont{S.}~\bibnamefont{More}}, \bibinfo{author}{\bibfnamefont{X.}~\bibnamefont{Li}}, \bibinfo{author}{\bibfnamefont{R.}~\bibnamefont{Dalal}}, \bibinfo{author}{\bibfnamefont{M.~M.} \bibnamefont{Rau}}, \bibinfo{author}{\bibfnamefont{J.}~\bibnamefont{Shi}}, \bibinfo{author}{\bibfnamefont{I.-N.} \bibnamefont{Chiu}}, \bibinfo{author}{\bibfnamefont{M.}~\bibnamefont{Shirasaki}}, \bibnamefont{et~al.}, \bibinfo{journal}{arXiv e-prints}  (\bibinfo{year}{2023}).
    
    \bibitem[{\citenamefont{White et~al.}(2010)\citenamefont{White, Cohn, and Smit}}]{whiteClusterGalaxyDynamics2010}
    \bibinfo{author}{\bibfnamefont{M.}~\bibnamefont{White}}, \bibinfo{author}{\bibfnamefont{J.~D.} \bibnamefont{Cohn}}, \bibnamefont{and} \bibinfo{author}{\bibfnamefont{R.}~\bibnamefont{Smit}}, \bibinfo{journal}{Monthly Notices of the Royal Astronomical Society} \textbf{\bibinfo{volume}{408}}, \bibinfo{pages}{1818} (\bibinfo{year}{2010}), ISSN \bibinfo{issn}{0035-8711}.
    
    \bibitem[{\citenamefont{Evrard et~al.}(2014)\citenamefont{Evrard, Arnault, Huterer, and Farahi}}]{evrardModelMultipropertyGalaxy2014}
    \bibinfo{author}{\bibfnamefont{A.~E.} \bibnamefont{Evrard}}, \bibinfo{author}{\bibfnamefont{P.}~\bibnamefont{Arnault}}, \bibinfo{author}{\bibfnamefont{D.}~\bibnamefont{Huterer}}, \bibnamefont{and} \bibinfo{author}{\bibfnamefont{A.}~\bibnamefont{Farahi}}, \bibinfo{journal}{Monthly Notices of the Royal Astronomical Society} \textbf{\bibinfo{volume}{441}}, \bibinfo{pages}{3562} (\bibinfo{year}{2014}), ISSN \bibinfo{issn}{1365-2966, 0035-8711}, \eprint{1403.1456}.
    
    \bibitem[{\citenamefont{{The Dark Energy Survey Collaboration}}(2005)}]{2005astro.ph.10346T}
    \bibinfo{author}{\bibnamefont{{The Dark Energy Survey Collaboration}}}, \bibinfo{journal}{arXiv e-prints} \bibinfo{eid}{astro-ph/0510346} (\bibinfo{year}{2005}), \eprint{astro-ph/0510346}.
    
    \bibitem[{\citenamefont{{LSST Science Collaboration} et~al.}(2009)\citenamefont{{LSST Science Collaboration}, {Abell}, {Allison}, {Anderson}, {Andrew}, {Angel}, {Armus}, {Arnett}, {Asztalos}, {Axelrod} et~al.}}]{2009arXiv0912.0201L}
    \bibinfo{author}{\bibnamefont{{LSST Science Collaboration}}}, \bibinfo{author}{\bibfnamefont{P.~A.} \bibnamefont{{Abell}}}, \bibinfo{author}{\bibfnamefont{J.}~\bibnamefont{{Allison}}}, \bibinfo{author}{\bibfnamefont{S.~F.} \bibnamefont{{Anderson}}}, \bibinfo{author}{\bibfnamefont{J.~R.} \bibnamefont{{Andrew}}}, \bibinfo{author}{\bibfnamefont{J.~R.~P.} \bibnamefont{{Angel}}}, \bibinfo{author}{\bibfnamefont{L.}~\bibnamefont{{Armus}}}, \bibinfo{author}{\bibfnamefont{D.}~\bibnamefont{{Arnett}}}, \bibinfo{author}{\bibfnamefont{S.~J.} \bibnamefont{{Asztalos}}}, \bibinfo{author}{\bibfnamefont{T.~S.} \bibnamefont{{Axelrod}}}, \bibnamefont{et~al.}, \bibinfo{journal}{arXiv e-prints} \bibinfo{eid}{arXiv:0912.0201} (\bibinfo{year}{2009}), \eprint{0912.0201}.
    
    \bibitem[{\citenamefont{{Laureijs} et~al.}(2011)\citenamefont{{Laureijs}, {Amiaux}, {Arduini}, {Augu{\`e}res}, {Brinchmann}, {Cole}, {Cropper}, {Dabin}, {Duvet}, {Ealet} et~al.}}]{2011arXiv1110.3193L}
    \bibinfo{author}{\bibfnamefont{R.}~\bibnamefont{{Laureijs}}}, \bibinfo{author}{\bibfnamefont{J.}~\bibnamefont{{Amiaux}}}, \bibinfo{author}{\bibfnamefont{S.}~\bibnamefont{{Arduini}}}, \bibinfo{author}{\bibfnamefont{J.~L.} \bibnamefont{{Augu{\`e}res}}}, \bibinfo{author}{\bibfnamefont{J.}~\bibnamefont{{Brinchmann}}}, \bibinfo{author}{\bibfnamefont{R.}~\bibnamefont{{Cole}}}, \bibinfo{author}{\bibfnamefont{M.}~\bibnamefont{{Cropper}}}, \bibinfo{author}{\bibfnamefont{C.}~\bibnamefont{{Dabin}}}, \bibinfo{author}{\bibfnamefont{L.}~\bibnamefont{{Duvet}}}, \bibinfo{author}{\bibfnamefont{A.}~\bibnamefont{{Ealet}}}, \bibnamefont{et~al.}, \bibinfo{journal}{arXiv e-prints} \bibinfo{eid}{arXiv:1110.3193} (\bibinfo{year}{2011}), \eprint{1110.3193}.
    
    \bibitem[{\citenamefont{{de Jong} et~al.}(2013)\citenamefont{{de Jong}, Verdoes~Kleijn, Kuijken, and Valentijn}}]{dejongKiloDegreeSurvey2013}
    \bibinfo{author}{\bibfnamefont{J.~T.~A.} \bibnamefont{{de Jong}}}, \bibinfo{author}{\bibfnamefont{G.~A.} \bibnamefont{Verdoes~Kleijn}}, \bibinfo{author}{\bibfnamefont{K.~H.} \bibnamefont{Kuijken}}, \bibnamefont{and} \bibinfo{author}{\bibfnamefont{E.~A.} \bibnamefont{Valentijn}}, \bibinfo{journal}{Exper. Astron.} \textbf{\bibinfo{volume}{35}}, \bibinfo{pages}{25} (\bibinfo{year}{2013}).
    
    \bibitem[{\citenamefont{Aihara et~al.}(2018)}]{aiharaHyperSuprimeCamSSP2018}
    \bibinfo{author}{\bibfnamefont{H.}~\bibnamefont{Aihara}} \bibnamefont{et~al.}, \bibinfo{journal}{Publ. Astron. Soc. Jap.} \textbf{\bibinfo{volume}{70}}, \bibinfo{pages}{S4} (\bibinfo{year}{2018}).
    
    \bibitem[{\citenamefont{{Dor{\'e}} et~al.}(2018)\citenamefont{{Dor{\'e}}, {Hirata}, {Wang}, {Weinberg}, {Baronchelli}, {Benson}, {Capak}, {Choi}, {Eifler}, {Hemmati} et~al.}}]{2018arXiv180403628D}
    \bibinfo{author}{\bibfnamefont{O.}~\bibnamefont{{Dor{\'e}}}}, \bibinfo{author}{\bibfnamefont{C.}~\bibnamefont{{Hirata}}}, \bibinfo{author}{\bibfnamefont{Y.}~\bibnamefont{{Wang}}}, \bibinfo{author}{\bibfnamefont{D.}~\bibnamefont{{Weinberg}}}, \bibinfo{author}{\bibfnamefont{I.}~\bibnamefont{{Baronchelli}}}, \bibinfo{author}{\bibfnamefont{A.}~\bibnamefont{{Benson}}}, \bibinfo{author}{\bibfnamefont{P.}~\bibnamefont{{Capak}}}, \bibinfo{author}{\bibfnamefont{A.}~\bibnamefont{{Choi}}}, \bibinfo{author}{\bibfnamefont{T.}~\bibnamefont{{Eifler}}}, \bibinfo{author}{\bibfnamefont{S.}~\bibnamefont{{Hemmati}}}, \bibnamefont{et~al.}, \bibinfo{journal}{arXiv e-prints} \bibinfo{eid}{arXiv:1804.03628} (\bibinfo{year}{2018}), \eprint{1804.03628}.
    
    \bibitem[{\citenamefont{DeRose et~al.}(2019)\citenamefont{DeRose, Wechsler, Becker, Busha, Rykoff, MacCrann, Erickson, Evrard, Kravtsov, Gruen et~al.}}]{deroseBuzzardFlockDark2019}
    \bibinfo{author}{\bibfnamefont{J.}~\bibnamefont{DeRose}}, \bibinfo{author}{\bibfnamefont{R.~H.} \bibnamefont{Wechsler}}, \bibinfo{author}{\bibfnamefont{M.~R.} \bibnamefont{Becker}}, \bibinfo{author}{\bibfnamefont{M.~T.} \bibnamefont{Busha}}, \bibinfo{author}{\bibfnamefont{E.~S.} \bibnamefont{Rykoff}}, \bibinfo{author}{\bibfnamefont{N.}~\bibnamefont{MacCrann}}, \bibinfo{author}{\bibfnamefont{B.}~\bibnamefont{Erickson}}, \bibinfo{author}{\bibfnamefont{A.~E.} \bibnamefont{Evrard}}, \bibinfo{author}{\bibfnamefont{A.}~\bibnamefont{Kravtsov}}, \bibinfo{author}{\bibfnamefont{D.}~\bibnamefont{Gruen}}, \bibnamefont{et~al.}, \bibinfo{journal}{arXiv:1901.02401 [astro-ph]}  (\bibinfo{year}{2019}), \eprint{1901.02401}.
    
    \bibitem[{\citenamefont{Salcedo et~al.}(2023)\citenamefont{Salcedo, Wu, Rozo, Weinberg, To, Sunayama, and Lee}}]{salcedoDarkEnergySurvey2023}
    \bibinfo{author}{\bibfnamefont{A.~N.} \bibnamefont{Salcedo}}, \bibinfo{author}{\bibfnamefont{H.-Y.} \bibnamefont{Wu}}, \bibinfo{author}{\bibfnamefont{E.}~\bibnamefont{Rozo}}, \bibinfo{author}{\bibfnamefont{D.~H.} \bibnamefont{Weinberg}}, \bibinfo{author}{\bibfnamefont{C.-H.} \bibnamefont{To}}, \bibinfo{author}{\bibfnamefont{T.}~\bibnamefont{Sunayama}}, \bibnamefont{and} \bibinfo{author}{\bibfnamefont{A.}~\bibnamefont{Lee}}, \bibinfo{journal}{arXiv e-prints}  (\bibinfo{year}{2023}).
    
    \bibitem[{\citenamefont{Zeng et~al.}(2023)\citenamefont{Zeng, Salcedo, Wu, and Hirata}}]{zengSelfcalibratingOpticalGalaxy2023}
    \bibinfo{author}{\bibfnamefont{C.}~\bibnamefont{Zeng}}, \bibinfo{author}{\bibfnamefont{A.~N.} \bibnamefont{Salcedo}}, \bibinfo{author}{\bibfnamefont{H.-Y.} \bibnamefont{Wu}}, \bibnamefont{and} \bibinfo{author}{\bibfnamefont{C.~M.} \bibnamefont{Hirata}}, \bibinfo{journal}{Monthly Notices of the Royal Astronomical Society} \textbf{\bibinfo{volume}{523}}, \bibinfo{pages}{4270} (\bibinfo{year}{2023}), ISSN \bibinfo{issn}{0035-8711}.
    
    \bibitem[{\citenamefont{Berlind and Weinberg}(2002)}]{berlindHaloOccupationDistribution2002}
    \bibinfo{author}{\bibfnamefont{A.~A.} \bibnamefont{Berlind}} \bibnamefont{and} \bibinfo{author}{\bibfnamefont{D.~H.} \bibnamefont{Weinberg}}, \bibinfo{journal}{The Astrophysical Journal} \textbf{\bibinfo{volume}{575}}, \bibinfo{pages}{587} (\bibinfo{year}{2002}), ISSN \bibinfo{issn}{0004-637X}.
    
    \bibitem[{\citenamefont{Kravtsov et~al.}(2004)\citenamefont{Kravtsov, Berlind, Wechsler, Klypin, Gottl{\"o}ber, Allgood, and Primack}}]{kravtsovDarkSideHalo2004}
    \bibinfo{author}{\bibfnamefont{A.~V.} \bibnamefont{Kravtsov}}, \bibinfo{author}{\bibfnamefont{A.~A.} \bibnamefont{Berlind}}, \bibinfo{author}{\bibfnamefont{R.~H.} \bibnamefont{Wechsler}}, \bibinfo{author}{\bibfnamefont{A.~A.} \bibnamefont{Klypin}}, \bibinfo{author}{\bibfnamefont{S.}~\bibnamefont{Gottl{\"o}ber}}, \bibinfo{author}{\bibfnamefont{B.}~\bibnamefont{Allgood}}, \bibnamefont{and} \bibinfo{author}{\bibfnamefont{J.~R.} \bibnamefont{Primack}}, \bibinfo{journal}{The Astrophysical Journal} \textbf{\bibinfo{volume}{609}}, \bibinfo{pages}{35} (\bibinfo{year}{2004}), ISSN \bibinfo{issn}{0004-637X}.
    
    \bibitem[{\citenamefont{Zheng et~al.}(2005)\citenamefont{Zheng, Berlind, Weinberg, Benson, Baugh, Cole, Dav{\'e}, Frenk, Katz, and Lacey}}]{zhengTheoreticalModelsHalo2005}
    \bibinfo{author}{\bibfnamefont{Z.}~\bibnamefont{Zheng}}, \bibinfo{author}{\bibfnamefont{A.~A.} \bibnamefont{Berlind}}, \bibinfo{author}{\bibfnamefont{D.~H.} \bibnamefont{Weinberg}}, \bibinfo{author}{\bibfnamefont{A.~J.} \bibnamefont{Benson}}, \bibinfo{author}{\bibfnamefont{C.~M.} \bibnamefont{Baugh}}, \bibinfo{author}{\bibfnamefont{S.}~\bibnamefont{Cole}}, \bibinfo{author}{\bibfnamefont{R.}~\bibnamefont{Dav{\'e}}}, \bibinfo{author}{\bibfnamefont{C.~S.} \bibnamefont{Frenk}}, \bibinfo{author}{\bibfnamefont{N.}~\bibnamefont{Katz}}, \bibnamefont{and} \bibinfo{author}{\bibfnamefont{C.~G.} \bibnamefont{Lacey}}, \bibinfo{journal}{The Astrophysical Journal} \textbf{\bibinfo{volume}{633}}, \bibinfo{pages}{791} (\bibinfo{year}{2005}), ISSN \bibinfo{issn}{0004-637X}.
    
    \bibitem[{\citenamefont{Costanzi et~al.}(2019)\citenamefont{Costanzi, Rozo, Rykoff, Farahi, Jeltema, Evrard, Mantz, Gruen, Mandelbaum, DeRose et~al.}}]{costanziModelingProjectionEffects2019}
    \bibinfo{author}{\bibfnamefont{M.}~\bibnamefont{Costanzi}}, \bibinfo{author}{\bibfnamefont{E.}~\bibnamefont{Rozo}}, \bibinfo{author}{\bibfnamefont{E.~S.} \bibnamefont{Rykoff}}, \bibinfo{author}{\bibfnamefont{A.}~\bibnamefont{Farahi}}, \bibinfo{author}{\bibfnamefont{T.}~\bibnamefont{Jeltema}}, \bibinfo{author}{\bibfnamefont{A.~E.} \bibnamefont{Evrard}}, \bibinfo{author}{\bibfnamefont{A.}~\bibnamefont{Mantz}}, \bibinfo{author}{\bibfnamefont{D.}~\bibnamefont{Gruen}}, \bibinfo{author}{\bibfnamefont{R.}~\bibnamefont{Mandelbaum}}, \bibinfo{author}{\bibfnamefont{J.}~\bibnamefont{DeRose}}, \bibnamefont{et~al.}, \bibinfo{journal}{Monthly Notices of the Royal Astronomical Society} \textbf{\bibinfo{volume}{482}}, \bibinfo{pages}{490} (\bibinfo{year}{2019}), ISSN \bibinfo{issn}{0035-8711, 1365-2966}, \eprint{1807.07072}.
    
    \bibitem[{\citenamefont{Tinker et~al.}(2008)\citenamefont{Tinker, Kravtsov, Klypin, Abazajian, Warren, Yepes, Gottlober, and Holz}}]{tinkerHaloMassFunction2008}
    \bibinfo{author}{\bibfnamefont{J.~L.} \bibnamefont{Tinker}}, \bibinfo{author}{\bibfnamefont{A.~V.} \bibnamefont{Kravtsov}}, \bibinfo{author}{\bibfnamefont{A.}~\bibnamefont{Klypin}}, \bibinfo{author}{\bibfnamefont{K.}~\bibnamefont{Abazajian}}, \bibinfo{author}{\bibfnamefont{M.~S.} \bibnamefont{Warren}}, \bibinfo{author}{\bibfnamefont{G.}~\bibnamefont{Yepes}}, \bibinfo{author}{\bibfnamefont{S.}~\bibnamefont{Gottlober}}, \bibnamefont{and} \bibinfo{author}{\bibfnamefont{D.~E.} \bibnamefont{Holz}}, \bibinfo{journal}{The Astrophysical Journal} \textbf{\bibinfo{volume}{688}}, \bibinfo{pages}{709} (\bibinfo{year}{2008}), ISSN \bibinfo{issn}{0004-637X, 1538-4357}, \eprint{0803.2706}.
    
    \bibitem[{\citenamefont{Nishimichi et~al.}(2019)\citenamefont{Nishimichi, Takada, Takahashi, Osato, Shirasaki, Oogi, Miyatake, Oguri, Murata, Kobayashi et~al.}}]{nishimichiDarkQuestFast2019}
    \bibinfo{author}{\bibfnamefont{T.}~\bibnamefont{Nishimichi}}, \bibinfo{author}{\bibfnamefont{M.}~\bibnamefont{Takada}}, \bibinfo{author}{\bibfnamefont{R.}~\bibnamefont{Takahashi}}, \bibinfo{author}{\bibfnamefont{K.}~\bibnamefont{Osato}}, \bibinfo{author}{\bibfnamefont{M.}~\bibnamefont{Shirasaki}}, \bibinfo{author}{\bibfnamefont{T.}~\bibnamefont{Oogi}}, \bibinfo{author}{\bibfnamefont{H.}~\bibnamefont{Miyatake}}, \bibinfo{author}{\bibfnamefont{M.}~\bibnamefont{Oguri}}, \bibinfo{author}{\bibfnamefont{R.}~\bibnamefont{Murata}}, \bibinfo{author}{\bibfnamefont{Y.}~\bibnamefont{Kobayashi}}, \bibnamefont{et~al.}, \bibinfo{journal}{The Astrophysical Journal} \textbf{\bibinfo{volume}{884}}, \bibinfo{pages}{29} (\bibinfo{year}{2019}), ISSN \bibinfo{issn}{1538-4357}, \eprint{1811.09504}.
    
    \bibitem[{\citenamefont{To et~al.}(2023)\citenamefont{To, DeRose, Wechsler, Rykoff, Wu, Adhikari, Krause, Rozo, and Weinberg}}]{toBuzzardCardinalImproved2023a}
    \bibinfo{author}{\bibfnamefont{C.-H.} \bibnamefont{To}}, \bibinfo{author}{\bibfnamefont{J.}~\bibnamefont{DeRose}}, \bibinfo{author}{\bibfnamefont{R.~H.} \bibnamefont{Wechsler}}, \bibinfo{author}{\bibfnamefont{E.}~\bibnamefont{Rykoff}}, \bibinfo{author}{\bibfnamefont{H.-Y.} \bibnamefont{Wu}}, \bibinfo{author}{\bibfnamefont{S.}~\bibnamefont{Adhikari}}, \bibinfo{author}{\bibfnamefont{E.}~\bibnamefont{Krause}}, \bibinfo{author}{\bibfnamefont{E.}~\bibnamefont{Rozo}}, \bibnamefont{and} \bibinfo{author}{\bibfnamefont{D.~H.} \bibnamefont{Weinberg}}, \bibinfo{journal}{arXiv e-prints}  (\bibinfo{year}{2023}).
    
    \bibitem[{\citenamefont{Vanderlinde et~al.}(2010)\citenamefont{Vanderlinde, Crawford, {de Haan}, Dudley, Shaw, Ade, Aird, Benson, Bleem, Brodwin et~al.}}]{vanderlindeGalaxyClustersSelected2010}
    \bibinfo{author}{\bibfnamefont{K.}~\bibnamefont{Vanderlinde}}, \bibinfo{author}{\bibfnamefont{T.~M.} \bibnamefont{Crawford}}, \bibinfo{author}{\bibfnamefont{T.}~\bibnamefont{{de Haan}}}, \bibinfo{author}{\bibfnamefont{J.~P.} \bibnamefont{Dudley}}, \bibinfo{author}{\bibfnamefont{L.}~\bibnamefont{Shaw}}, \bibinfo{author}{\bibfnamefont{P.~A.~R.} \bibnamefont{Ade}}, \bibinfo{author}{\bibfnamefont{K.~A.} \bibnamefont{Aird}}, \bibinfo{author}{\bibfnamefont{B.~A.} \bibnamefont{Benson}}, \bibinfo{author}{\bibfnamefont{L.~E.} \bibnamefont{Bleem}}, \bibinfo{author}{\bibfnamefont{M.}~\bibnamefont{Brodwin}}, \bibnamefont{et~al.}, \bibinfo{journal}{The Astrophysical Journal} \textbf{\bibinfo{volume}{722}}, \bibinfo{pages}{1180} (\bibinfo{year}{2010}), ISSN \bibinfo{issn}{0004-637X, 1538-4357}, \eprint{1003.0003}.
    
    \bibitem[{\citenamefont{Garrison et~al.}(2018)\citenamefont{Garrison, Eisenstein, Ferrer, Tinker, Pinto, and Weinberg}}]{garrisonAbacusCosmosSuite2018}
    \bibinfo{author}{\bibfnamefont{L.~H.} \bibnamefont{Garrison}}, \bibinfo{author}{\bibfnamefont{D.~J.} \bibnamefont{Eisenstein}}, \bibinfo{author}{\bibfnamefont{D.}~\bibnamefont{Ferrer}}, \bibinfo{author}{\bibfnamefont{J.~L.} \bibnamefont{Tinker}}, \bibinfo{author}{\bibfnamefont{P.~A.} \bibnamefont{Pinto}}, \bibnamefont{and} \bibinfo{author}{\bibfnamefont{D.~H.} \bibnamefont{Weinberg}}, \bibinfo{journal}{The Astrophysical Journal Supplement Series} \textbf{\bibinfo{volume}{236}}, \bibinfo{pages}{43} (\bibinfo{year}{2018}), ISSN \bibinfo{issn}{1538-4365}, \eprint{1712.05768}.
    
    \bibitem[{\citenamefont{Ade et~al.}(2016)\citenamefont{Ade, Aghanim, Arnaud, Ashdown, Aumont, Baccigalupi, Banday, Barreiro, Bartlett, Bartolo et~al.}}]{adePlanck2015Results2016}
    \bibinfo{author}{\bibfnamefont{P.~a.~R.} \bibnamefont{Ade}}, \bibinfo{author}{\bibfnamefont{N.}~\bibnamefont{Aghanim}}, \bibinfo{author}{\bibfnamefont{M.}~\bibnamefont{Arnaud}}, \bibinfo{author}{\bibfnamefont{M.}~\bibnamefont{Ashdown}}, \bibinfo{author}{\bibfnamefont{J.}~\bibnamefont{Aumont}}, \bibinfo{author}{\bibfnamefont{C.}~\bibnamefont{Baccigalupi}}, \bibinfo{author}{\bibfnamefont{A.~J.} \bibnamefont{Banday}}, \bibinfo{author}{\bibfnamefont{R.~B.} \bibnamefont{Barreiro}}, \bibinfo{author}{\bibfnamefont{J.~G.} \bibnamefont{Bartlett}}, \bibinfo{author}{\bibfnamefont{N.}~\bibnamefont{Bartolo}}, \bibnamefont{et~al.}, \bibinfo{journal}{Astronomy \& Astrophysics} \textbf{\bibinfo{volume}{594}}, \bibinfo{pages}{A13} (\bibinfo{year}{2016}), ISSN \bibinfo{issn}{0004-6361, 1432-0746}.
    
    \bibitem[{\citenamefont{Behroozi et~al.}(2012)\citenamefont{Behroozi, Wechsler, and Wu}}]{behrooziROCKSTARPHASESPACETEMPORAL2012}
    \bibinfo{author}{\bibfnamefont{P.~S.} \bibnamefont{Behroozi}}, \bibinfo{author}{\bibfnamefont{R.~H.} \bibnamefont{Wechsler}}, \bibnamefont{and} \bibinfo{author}{\bibfnamefont{H.-Y.} \bibnamefont{Wu}}, \bibinfo{journal}{The Astrophysical Journal} \textbf{\bibinfo{volume}{762}}, \bibinfo{pages}{109} (\bibinfo{year}{2012}), ISSN \bibinfo{issn}{0004-637X}.
    
    \bibitem[{\citenamefont{{Salcedo} et~al.}(2023)\citenamefont{{Salcedo}, {Wu}, {Rozo}, {Weinberg}, {To}, {Sunayama}, and {Lee}}}]{2023arXiv231003944S}
    \bibinfo{author}{\bibfnamefont{A.~N.} \bibnamefont{{Salcedo}}}, \bibinfo{author}{\bibfnamefont{H.-Y.} \bibnamefont{{Wu}}}, \bibinfo{author}{\bibfnamefont{E.}~\bibnamefont{{Rozo}}}, \bibinfo{author}{\bibfnamefont{D.~H.} \bibnamefont{{Weinberg}}}, \bibinfo{author}{\bibfnamefont{C.-H.} \bibnamefont{{To}}}, \bibinfo{author}{\bibfnamefont{T.}~\bibnamefont{{Sunayama}}}, \bibnamefont{and} \bibinfo{author}{\bibfnamefont{A.}~\bibnamefont{{Lee}}}, \bibinfo{journal}{arXiv e-prints} \bibinfo{eid}{arXiv:2310.03944} (\bibinfo{year}{2023}), \eprint{2310.03944}.
    
    \bibitem[{\citenamefont{Sinha and Garrison}(2020)}]{sinhaCORRFUNCSuiteBlazing2020}
    \bibinfo{author}{\bibfnamefont{M.}~\bibnamefont{Sinha}} \bibnamefont{and} \bibinfo{author}{\bibfnamefont{L.~H.} \bibnamefont{Garrison}}, \bibinfo{journal}{Monthly Notices of the Royal Astronomical Society} \textbf{\bibinfo{volume}{491}}, \bibinfo{pages}{3022} (\bibinfo{year}{2020}), ISSN \bibinfo{issn}{0035-8711}.
    
    \bibitem[{\citenamefont{Zuntz et~al.}(2018)\citenamefont{Zuntz, Sheldon, Samuroff, Troxel, Jarvis, MacCrann, Gruen, Prat, S{\'a}nchez, Choi et~al.}}]{zuntzDarkEnergySurvey2018}
    \bibinfo{author}{\bibfnamefont{J.}~\bibnamefont{Zuntz}}, \bibinfo{author}{\bibfnamefont{E.}~\bibnamefont{Sheldon}}, \bibinfo{author}{\bibfnamefont{S.}~\bibnamefont{Samuroff}}, \bibinfo{author}{\bibfnamefont{M.~A.} \bibnamefont{Troxel}}, \bibinfo{author}{\bibfnamefont{M.}~\bibnamefont{Jarvis}}, \bibinfo{author}{\bibfnamefont{N.}~\bibnamefont{MacCrann}}, \bibinfo{author}{\bibfnamefont{D.}~\bibnamefont{Gruen}}, \bibinfo{author}{\bibfnamefont{J.}~\bibnamefont{Prat}}, \bibinfo{author}{\bibfnamefont{C.}~\bibnamefont{S{\'a}nchez}}, \bibinfo{author}{\bibfnamefont{A.}~\bibnamefont{Choi}}, \bibnamefont{et~al.}, \bibinfo{journal}{Monthly Notices of the Royal Astronomical Society} \textbf{\bibinfo{volume}{481}}, \bibinfo{pages}{1149} (\bibinfo{year}{2018}), ISSN \bibinfo{issn}{0035-8711}.
    
    \bibitem[{\citenamefont{{Foreman-Mackey} et~al.}(2013)\citenamefont{{Foreman-Mackey}, Hogg, Lang, and Goodman}}]{foreman-mackeyEmceeMCMCHammer2013}
    \bibinfo{author}{\bibfnamefont{D.}~\bibnamefont{{Foreman-Mackey}}}, \bibinfo{author}{\bibfnamefont{D.~W.} \bibnamefont{Hogg}}, \bibinfo{author}{\bibfnamefont{D.}~\bibnamefont{Lang}}, \bibnamefont{and} \bibinfo{author}{\bibfnamefont{J.}~\bibnamefont{Goodman}}, \bibinfo{journal}{Publications of the Astronomical Society of the Pacific} \textbf{\bibinfo{volume}{125}}, \bibinfo{pages}{306} (\bibinfo{year}{2013}), ISSN \bibinfo{issn}{00046280, 15383873}, \eprint{1202.3665}.
    
    \bibitem[{\citenamefont{Goodman and Weare}(2010)}]{goodmanEnsembleSamplersAffine2010}
    \bibinfo{author}{\bibfnamefont{J.}~\bibnamefont{Goodman}} \bibnamefont{and} \bibinfo{author}{\bibfnamefont{J.}~\bibnamefont{Weare}}, \bibinfo{journal}{Communications in Applied Mathematics and Computational Science} \textbf{\bibinfo{volume}{5}}, \bibinfo{pages}{65} (\bibinfo{year}{2010}), ISSN \bibinfo{issn}{2157-5452}.
    
    \bibitem[{\citenamefont{Wu et~al.}(2021)\citenamefont{Wu, Weinberg, Salcedo, and Wibking}}]{wuCosmologyGalaxyCluster2021}
    \bibinfo{author}{\bibfnamefont{H.-Y.} \bibnamefont{Wu}}, \bibinfo{author}{\bibfnamefont{D.~H.} \bibnamefont{Weinberg}}, \bibinfo{author}{\bibfnamefont{A.~N.} \bibnamefont{Salcedo}}, \bibnamefont{and} \bibinfo{author}{\bibfnamefont{B.~D.} \bibnamefont{Wibking}}, \bibinfo{journal}{The Astrophysical Journal} \textbf{\bibinfo{volume}{910}}, \bibinfo{pages}{28} (\bibinfo{year}{2021}), ISSN \bibinfo{issn}{0004-637X, 1538-4357}, \eprint{2012.01956}.
    
    \bibitem[{\citenamefont{Rozo et~al.}(2015)\citenamefont{Rozo, Rykoff, Bartlett, and Melin}}]{rozoRedMaPPerIIIDetailed2015}
    \bibinfo{author}{\bibfnamefont{E.}~\bibnamefont{Rozo}}, \bibinfo{author}{\bibfnamefont{E.~S.} \bibnamefont{Rykoff}}, \bibinfo{author}{\bibfnamefont{J.~G.} \bibnamefont{Bartlett}}, \bibnamefont{and} \bibinfo{author}{\bibfnamefont{J.~B.} \bibnamefont{Melin}}, \bibinfo{journal}{Monthly Notices of the Royal Astronomical Society} \textbf{\bibinfo{volume}{450}}, \bibinfo{pages}{592} (\bibinfo{year}{2015}), ISSN \bibinfo{issn}{0035-8711, 1365-2966}, \eprint{1401.7716}.
    
    \bibitem[{\citenamefont{Osato and Nagai}(2022)}]{osatoBaryonPastingAlgorithm2022b}
    \bibinfo{author}{\bibfnamefont{K.}~\bibnamefont{Osato}} \bibnamefont{and} \bibinfo{author}{\bibfnamefont{D.}~\bibnamefont{Nagai}}, \bibinfo{journal}{Monthly Notices of the Royal Astronomical Society} \textbf{\bibinfo{volume}{519}}, \bibinfo{pages}{2069} (\bibinfo{year}{2022}), ISSN \bibinfo{issn}{0035-8711, 1365-2966}, \eprint{2201.02632}.
    
    \bibitem[{\citenamefont{Henson et~al.}(2017)\citenamefont{Henson, Barnes, Kay, McCarthy, and Schaye}}]{hensonImpactBaryonsMassive2017}
    \bibinfo{author}{\bibfnamefont{M.~A.} \bibnamefont{Henson}}, \bibinfo{author}{\bibfnamefont{D.~J.} \bibnamefont{Barnes}}, \bibinfo{author}{\bibfnamefont{S.~T.} \bibnamefont{Kay}}, \bibinfo{author}{\bibfnamefont{I.~G.} \bibnamefont{McCarthy}}, \bibnamefont{and} \bibinfo{author}{\bibfnamefont{J.}~\bibnamefont{Schaye}}, \bibinfo{journal}{Monthly Notices of the Royal Astronomical Society} \textbf{\bibinfo{volume}{465}}, \bibinfo{pages}{3361} (\bibinfo{year}{2017}), ISSN \bibinfo{issn}{0035-8711}.
    
    \bibitem[{\citenamefont{Debackere et~al.}(2021)\citenamefont{Debackere, Schaye, and Hoekstra}}]{debackereHowBaryonsCan2021}
    \bibinfo{author}{\bibfnamefont{S.~N.~B.} \bibnamefont{Debackere}}, \bibinfo{author}{\bibfnamefont{J.}~\bibnamefont{Schaye}}, \bibnamefont{and} \bibinfo{author}{\bibfnamefont{H.}~\bibnamefont{Hoekstra}}, \bibinfo{journal}{Monthly Notices of the Royal Astronomical Society} \textbf{\bibinfo{volume}{505}}, \bibinfo{pages}{593} (\bibinfo{year}{2021}), ISSN \bibinfo{issn}{0035-8711, 1365-2966}, \eprint{2101.07800}.
    
    \bibitem[{\citenamefont{Giri and Schneider}(2021)}]{giriEmulationBaryonicEffects2021}
    \bibinfo{author}{\bibfnamefont{S.~K.} \bibnamefont{Giri}} \bibnamefont{and} \bibinfo{author}{\bibfnamefont{A.}~\bibnamefont{Schneider}}, \bibinfo{journal}{Journal of Cosmology and Astroparticle Physics} \textbf{\bibinfo{volume}{2021}}, \bibinfo{pages}{046} (\bibinfo{year}{2021}), ISSN \bibinfo{issn}{1475-7516}, \eprint{2108.08863}.
    
    \bibitem[{\citenamefont{{Grandis} et~al.}(2023)\citenamefont{{Grandis}, {Arico'}, {Schneider}, and {Linke}}}]{2023arXiv230902920G}
    \bibinfo{author}{\bibfnamefont{S.}~\bibnamefont{{Grandis}}}, \bibinfo{author}{\bibfnamefont{G.}~\bibnamefont{{Arico'}}}, \bibinfo{author}{\bibfnamefont{A.}~\bibnamefont{{Schneider}}}, \bibnamefont{and} \bibinfo{author}{\bibfnamefont{L.}~\bibnamefont{{Linke}}}, \bibinfo{journal}{arXiv e-prints} \bibinfo{eid}{arXiv:2309.02920} (\bibinfo{year}{2023}), \eprint{2309.02920}.
    
    \bibitem[{\citenamefont{Zhang et~al.}(2019)\citenamefont{Zhang, Yanny, Palmese, Gruen, To, Rykoff, Leung, Collins, Hilton, Abbott et~al.}}]{zhangDarkEnergySurvey2019}
    \bibinfo{author}{\bibfnamefont{Y.}~\bibnamefont{Zhang}}, \bibinfo{author}{\bibfnamefont{B.}~\bibnamefont{Yanny}}, \bibinfo{author}{\bibfnamefont{A.}~\bibnamefont{Palmese}}, \bibinfo{author}{\bibfnamefont{D.}~\bibnamefont{Gruen}}, \bibinfo{author}{\bibfnamefont{C.}~\bibnamefont{To}}, \bibinfo{author}{\bibfnamefont{E.~S.} \bibnamefont{Rykoff}}, \bibinfo{author}{\bibfnamefont{Y.}~\bibnamefont{Leung}}, \bibinfo{author}{\bibfnamefont{C.}~\bibnamefont{Collins}}, \bibinfo{author}{\bibfnamefont{M.}~\bibnamefont{Hilton}}, \bibinfo{author}{\bibfnamefont{T.~M.~C.} \bibnamefont{Abbott}}, \bibnamefont{et~al.}, \bibinfo{journal}{The Astrophysical Journal} \textbf{\bibinfo{volume}{874}}, \bibinfo{pages}{165} (\bibinfo{year}{2019}), ISSN \bibinfo{issn}{1538-4357}, \eprint{1812.04004}.
    
    \bibitem[{\citenamefont{{Kelly} et~al.}(2023)\citenamefont{{Kelly}, {Jobel}, {Eiger}, {Abd}, {Jeltema}, {Giles}, {Hollowood}, {Wilkinson}, {Turner}, {Bhargava} et~al.}}]{2023arXiv231013207K}
    \bibinfo{author}{\bibfnamefont{P.}~\bibnamefont{{Kelly}}}, \bibinfo{author}{\bibfnamefont{J.}~\bibnamefont{{Jobel}}}, \bibinfo{author}{\bibfnamefont{O.}~\bibnamefont{{Eiger}}}, \bibinfo{author}{\bibfnamefont{A.}~\bibnamefont{{Abd}}}, \bibinfo{author}{\bibfnamefont{T.~E.} \bibnamefont{{Jeltema}}}, \bibinfo{author}{\bibfnamefont{P.}~\bibnamefont{{Giles}}}, \bibinfo{author}{\bibfnamefont{D.~L.} \bibnamefont{{Hollowood}}}, \bibinfo{author}{\bibfnamefont{R.~D.} \bibnamefont{{Wilkinson}}}, \bibinfo{author}{\bibfnamefont{D.~J.} \bibnamefont{{Turner}}}, \bibinfo{author}{\bibfnamefont{S.}~\bibnamefont{{Bhargava}}}, \bibnamefont{et~al.}, \bibinfo{journal}{arXiv e-prints} \bibinfo{eid}{arXiv:2310.13207} (\bibinfo{year}{2023}), \eprint{2310.13207}.
    
    \bibitem[{\citenamefont{Grandis et~al.}(2021)\citenamefont{Grandis, Bocquet, Mohr, Klein, and Dolag}}]{grandisCalibrationBiasScatter2021}
    \bibinfo{author}{\bibfnamefont{S.}~\bibnamefont{Grandis}}, \bibinfo{author}{\bibfnamefont{S.}~\bibnamefont{Bocquet}}, \bibinfo{author}{\bibfnamefont{J.~J.} \bibnamefont{Mohr}}, \bibinfo{author}{\bibfnamefont{M.}~\bibnamefont{Klein}}, \bibnamefont{and} \bibinfo{author}{\bibfnamefont{K.}~\bibnamefont{Dolag}}, \bibinfo{journal}{Monthly Notices of the Royal Astronomical Society} \textbf{\bibinfo{volume}{507}}, \bibinfo{pages}{5671} (\bibinfo{year}{2021}), ISSN \bibinfo{issn}{0035-8711, 1365-2966}, \eprint{2103.16212}.
    
    \bibitem[{\citenamefont{Heitmann et~al.}(2014)\citenamefont{Heitmann, Lawrence, Kwan, Habib, and Higdon}}]{heitmannCoyoteUniverseExtended2014}
    \bibinfo{author}{\bibfnamefont{K.}~\bibnamefont{Heitmann}}, \bibinfo{author}{\bibfnamefont{E.}~\bibnamefont{Lawrence}}, \bibinfo{author}{\bibfnamefont{J.}~\bibnamefont{Kwan}}, \bibinfo{author}{\bibfnamefont{S.}~\bibnamefont{Habib}}, \bibnamefont{and} \bibinfo{author}{\bibfnamefont{D.}~\bibnamefont{Higdon}}, \bibinfo{journal}{The Astrophysical Journal} \textbf{\bibinfo{volume}{780}}, \bibinfo{pages}{111} (\bibinfo{year}{2014}), ISSN \bibinfo{issn}{0004-637X}.
    
    \bibitem[{\citenamefont{Wibking et~al.}(2020)\citenamefont{Wibking, Weinberg, Salcedo, Wu, Singh, {Rodr{\'i}guez-Torres}, Garrison, and Eisenstein}}]{wibkingCosmologyGalaxygalaxyLensing2020}
    \bibinfo{author}{\bibfnamefont{B.~D.} \bibnamefont{Wibking}}, \bibinfo{author}{\bibfnamefont{D.~H.} \bibnamefont{Weinberg}}, \bibinfo{author}{\bibfnamefont{A.~N.} \bibnamefont{Salcedo}}, \bibinfo{author}{\bibfnamefont{H.-Y.} \bibnamefont{Wu}}, \bibinfo{author}{\bibfnamefont{S.}~\bibnamefont{Singh}}, \bibinfo{author}{\bibfnamefont{S.}~\bibnamefont{{Rodr{\'i}guez-Torres}}}, \bibinfo{author}{\bibfnamefont{L.~H.} \bibnamefont{Garrison}}, \bibnamefont{and} \bibinfo{author}{\bibfnamefont{D.~J.} \bibnamefont{Eisenstein}}, \bibinfo{journal}{Monthly Notices of the Royal Astronomical Society} \textbf{\bibinfo{volume}{492}}, \bibinfo{pages}{2872} (\bibinfo{year}{2020}), ISSN \bibinfo{issn}{0035-8711, 1365-2966}, \eprint{1907.06293}.
    
    \bibitem[{\citenamefont{Harris et~al.}(2020)\citenamefont{Harris, Millman, {van der Walt}, Gommers, Virtanen, Cournapeau, Wieser, Taylor, Berg, Smith et~al.}}]{harrisArrayProgrammingNumPy2020}
    \bibinfo{author}{\bibfnamefont{C.~R.} \bibnamefont{Harris}}, \bibinfo{author}{\bibfnamefont{K.~J.} \bibnamefont{Millman}}, \bibinfo{author}{\bibfnamefont{S.~J.} \bibnamefont{{van der Walt}}}, \bibinfo{author}{\bibfnamefont{R.}~\bibnamefont{Gommers}}, \bibinfo{author}{\bibfnamefont{P.}~\bibnamefont{Virtanen}}, \bibinfo{author}{\bibfnamefont{D.}~\bibnamefont{Cournapeau}}, \bibinfo{author}{\bibfnamefont{E.}~\bibnamefont{Wieser}}, \bibinfo{author}{\bibfnamefont{J.}~\bibnamefont{Taylor}}, \bibinfo{author}{\bibfnamefont{S.}~\bibnamefont{Berg}}, \bibinfo{author}{\bibfnamefont{N.~J.} \bibnamefont{Smith}}, \bibnamefont{et~al.}, \bibinfo{journal}{Nature} \textbf{\bibinfo{volume}{585}}, \bibinfo{pages}{357} (\bibinfo{year}{2020}), ISSN \bibinfo{issn}{1476-4687}.
    
    \bibitem[{\citenamefont{Diemer}(2018)}]{diemerCOLOSSUSPythonToolkit2018}
    \bibinfo{author}{\bibfnamefont{B.}~\bibnamefont{Diemer}}, \bibinfo{journal}{The Astrophysical Journal Supplement Series} \textbf{\bibinfo{volume}{239}}, \bibinfo{pages}{35} (\bibinfo{year}{2018}), ISSN \bibinfo{issn}{0067-0049}.
    
    \bibitem[{\citenamefont{Virtanen et~al.}(2020)\citenamefont{Virtanen, Gommers, Oliphant, Haberland, Reddy, Cournapeau, Burovski, Peterson, Weckesser, Bright et~al.}}]{virtanenSciPyFundamentalAlgorithms2020}
    \bibinfo{author}{\bibfnamefont{P.}~\bibnamefont{Virtanen}}, \bibinfo{author}{\bibfnamefont{R.}~\bibnamefont{Gommers}}, \bibinfo{author}{\bibfnamefont{T.~E.} \bibnamefont{Oliphant}}, \bibinfo{author}{\bibfnamefont{M.}~\bibnamefont{Haberland}}, \bibinfo{author}{\bibfnamefont{T.}~\bibnamefont{Reddy}}, \bibinfo{author}{\bibfnamefont{D.}~\bibnamefont{Cournapeau}}, \bibinfo{author}{\bibfnamefont{E.}~\bibnamefont{Burovski}}, \bibinfo{author}{\bibfnamefont{P.}~\bibnamefont{Peterson}}, \bibinfo{author}{\bibfnamefont{W.}~\bibnamefont{Weckesser}}, \bibinfo{author}{\bibfnamefont{J.}~\bibnamefont{Bright}}, \bibnamefont{et~al.}, \bibinfo{journal}{Nature Methods} \textbf{\bibinfo{volume}{17}}, \bibinfo{pages}{261} (\bibinfo{year}{2020}), ISSN \bibinfo{issn}{1548-7105}.
    
    \bibitem[{\citenamefont{{The Astropy Collaboration} et~al.}(2018)\citenamefont{{The Astropy Collaboration}, {Price-Whelan}, Sip{\H o}cz, G{\"u}nther, Lim, Crawford, Conseil, Shupe, Craig, Dencheva et~al.}}]{theastropycollaborationAstropyProjectBuilding2018}
    \bibinfo{author}{\bibnamefont{{The Astropy Collaboration}}}, \bibinfo{author}{\bibfnamefont{A.~M.} \bibnamefont{{Price-Whelan}}}, \bibinfo{author}{\bibfnamefont{B.~M.} \bibnamefont{Sip{\H o}cz}}, \bibinfo{author}{\bibfnamefont{H.~M.} \bibnamefont{G{\"u}nther}}, \bibinfo{author}{\bibfnamefont{P.~L.} \bibnamefont{Lim}}, \bibinfo{author}{\bibfnamefont{S.~M.} \bibnamefont{Crawford}}, \bibinfo{author}{\bibfnamefont{S.}~\bibnamefont{Conseil}}, \bibinfo{author}{\bibfnamefont{D.~L.} \bibnamefont{Shupe}}, \bibinfo{author}{\bibfnamefont{M.~W.} \bibnamefont{Craig}}, \bibinfo{author}{\bibfnamefont{N.}~\bibnamefont{Dencheva}}, \bibnamefont{et~al.}, \bibinfo{journal}{The Astronomical Journal} \textbf{\bibinfo{volume}{156}}, \bibinfo{pages}{123} (\bibinfo{year}{2018}), ISSN \bibinfo{issn}{1538-3881}.
    
    \bibitem[{\citenamefont{Granger and P{\'e}rez}(2021)}]{grangerJupyterThinkingStorytelling2021}
    \bibinfo{author}{\bibfnamefont{B.~E.} \bibnamefont{Granger}} \bibnamefont{and} \bibinfo{author}{\bibfnamefont{F.}~\bibnamefont{P{\'e}rez}}, \bibinfo{journal}{Computing in Science \& Engineering} \textbf{\bibinfo{volume}{23}}, \bibinfo{pages}{7} (\bibinfo{year}{2021}), ISSN \bibinfo{issn}{1558-366X}.
    
    \bibitem[{\citenamefont{Hunter}(2007)}]{hunterMatplotlib2DGraphics2007}
    \bibinfo{author}{\bibfnamefont{J.~D.} \bibnamefont{Hunter}}, \bibinfo{journal}{Computing in Science \& Engineering} \textbf{\bibinfo{volume}{9}}, \bibinfo{pages}{90} (\bibinfo{year}{2007}), ISSN \bibinfo{issn}{1558-366X}.
    
    \bibitem[{\citenamefont{Mantz}(2019)}]{mantzCopingSelectionEffects2019}
    \bibinfo{author}{\bibfnamefont{A.~B.} \bibnamefont{Mantz}}, \bibinfo{journal}{Monthly Notices of the Royal Astronomical Society} \textbf{\bibinfo{volume}{485}}, \bibinfo{pages}{4863} (\bibinfo{year}{2019}), ISSN \bibinfo{issn}{0035-8711}.
    
    \end{thebibliography}

\end{document}